# Tuning topological spin textures in size-tailored chiral magnet insulator particles

*Priya R. Baral*[a,b], *Victor Ukleev*[c], *Thomas LaGrange*[d], *Robert Cubitt*[e], *Ivica Živković*[f], *Henrik M. Rønnow*[f], *Jonathan S. White*[c], *Arnaud Magrez*[a,*]

a) Crystal Growth Facility, Institute of Physics, École Polytechnique Fédérale de Lausanne (EPFL), CH-1015 Lausanne, Switzerland

b) Chair of Computational Condensed Matter Physics, Institute of Physics, École Polytechnique Fédérale de Lausanne (EPFL), CH-1015 Lausanne, Switzerland

c) Laboratory for Neutron Scattering and Imaging (LNS), Paul Scherrer Institut (PSI), CH-5232 Villigen, Switzerland

d) Laboratory for Ultrafast Microscopy and Electron Scattering (LUMES), Institute of Physics, École Polytechnique Fédérale de Lausanne (EPFL), CH-1015 Lausanne, Switzerland

e) Institut Laue Langevin, Large Scale Structures, 71 Avenue des Martyrs CS 20156, 38042 Grenoble, France

f) Laboratory for Quantum Magnetism, Institute of Physics, École Polytechnique Fédérale de Lausanne (EPFL), CH-1015 Lausanne, Switzerland

* Email : arnaud.magrez@epfl.ch

**ABSTRACT**

Topological spin textures such as skyrmions hold high potential for use as magnetically active elements in diverse near-future applications. While skyrmions in metallic multilayers attract great attention in this context, unleashing the myriad potential of skyrmions for various applications requires the discovery and customization of alternative host system paradigms. Here we developed and applied a chemical method to synthesize octahedral particles of the chiral insulating skyrmion host $Cu_2OSeO_3$ with both narrow size distribution, and tailored dimensions approaching the nanoscale. Combining magnetometry and neutron scattering experiments with micromagnetic simulations, we show that the bulk phase diagram of $Cu_2OSeO_3$ changes dramatically below octahedral heights of 400 nm. Further particle size-dependent regimes are identified where various topological spin textures such as skyrmions, merons and bobbers can stabilize, prior to a lower critical octahedral height of ~190 nm below which no topological spin texture is found stable. These findings suggest conditions under which sparse topological spin textures confined to chiral magnet nanoparticles can be stable, and provide fresh potential for insulator-based application paradigms.

**INTRODUCTION**

In condensed matter physics, elucidation of the physical properties of novel materials in low dimensions is crucial from the viewpoint of fundamental physics as well as novel applications such as imaging, catalysis and information devices, amongst others.[1-6] In these materials, the reduction of at least one of the physical dimensions below a certain critical length results in geometrical confinement effects. The effect of surface symmetry breaking, due to an absence of neighboring atoms or the presence of defects, together with geometrically confined magnetic spin textures may give rise to other exotic phenomena and different physics as compared to the bulk single crystals.[7-9] A fertile ground to discover such unusual modifications are low dimensional samples of compounds hosting topological, finite-sized spin textures, the so called skyrmions.[10]

Magnetic skyrmions are topologically-protected multi-spin objects that can have a size between a few to hundreds of nanometers,[11] and which are stable in both metals and insulators alike.[12-16] Amongst the very few known insulator skyrmions hosts,[15,17] $Cu_2OSeO_3$ is the archetype with a chiral cubic crystal structure. In the bulk, the skyrmions form a triangular skyrmion lattice (SkL) that is stable over a region that extends over a few Kelvin directly below the magnetic ordering temperature $T_c$ ~ 58 K. Below $T_c$ the system further displays magnetoelectric (ME) coupling due to simultaneously broken space- and time-reversal symmetries.[18,19] Owing to ME coupling, skyrmion manipulation by electric field (EF) has been shown experimentally,[20-22] including EF-driven skyrmion creation and annihilation.[23,24]

At the same time, the aforementioned geometrical confinement effects in regular-shaped samples with dimensions approaching the nanoscale appears to modify the magnetic phase diagrams of chiral magnets drastically.[25-27] Skyrmions have been imaged while confined in FIB-shaped nanowires and nanodisks of the itinerant B20 chiral magnets MnSi or FeGe.[28-33] It is found that the phase diagram of MnSi nanowires differs from those of both thin plate and bulk samples, thus illustrating an effect of particle morphology though there is no report of the critical nanowire dimensions required for skyrmion formation. For insulating $Cu_2OSeO_3$, the hitherto most relevant study was performed on nanoparticles of average size ~40 nm, though the effects of confinement are complicated to identify due to the broad-size distribution of the synthesized nanoparticles.[26] Moreover, the average size lies below the size of a single skyrmion (60 nm), the latter being the crucial length-scale where the effects of geometrical confinement are expected to be most prominent. To date, the systematic study of the particle size-dependent phase diagram of skyrmion hosting chiral magnets remains missing, but such studies are essential for both understanding and controlling the emergent physics in these systems as all dimensions approach the nanoscale.

In the present work, we developed chemical approaches to achieve the large-scale synthesis of single crystal octahedral $Cu_2OSeO_3$ particles for the first time, and crucially, the precise control of the particle size. Both experiment and micromagnetic simulations reveal that an array of topological magnetic textures can become stable in the resulting particles, dependent on the octahedral height, $h$. For $h$ in the range of a few hundred nm, spiral and skyrmion spin textures can be stabilized, while the latter become unstable for $h$ below a critical value of 190 nm. These observations are consistent with the simulations that further implicate ranges of $h$ where other topological textures such as chiral bobbers and merons can also be stable and even co-exist with skyrmions. Our results obtained at low temperatures in $Cu_2OSeO_3$ particles provide insights that will be important for room temperature insulating skyrmionic nanomaterials discovered in the future, and their implementation in low power consumption non-volatile technologies such as magnonics applications.[23]

## MATERIALS AND METHODS

**Material preparation and structural characterizations.** The synthesis of $Cu_2OSeO_3$ particles proceeds during a low temperature reflux treatment performed in a round bottom flask at temperature ranging from 70 °C to 100 °C (Fig. 1A). The solution is kept under constant stirring for homogeneous distribution of particles and temperature throughout. $CuSeO_3 \cdot 2H_2O$, used as selenite precursor, is obtained by instantaneous precipitation when highly concentrated $CuSO_4 \cdot 5H_2O$ and $SeO_2$ solutions are mixed. Bases used during reflux are $NH_4OH$ or $Cu(OH)_2$. As shown in Fig.1 and 2, nanoparticles with a size smaller than 100 nm are obtained with $NH_4OH$ while the use of $Cu(OH)_2$ yields particles larger than 100 nm. $Cu(OH)_2$ is prepared by reaction between $Cu(NO_3)_2 \cdot 3H_2O$, $NH_4OH$ and $NaOH$.[34] Using high temporal resolution synchrotron X-ray diffraction (XRD), the transformation of $CuSeO_3 \cdot 2H_2O$ into $Cu_2OSeO_3$ can be followed *in situ*, as shown in Fig. S1A. The low temperatures ensure restrictive growth of the nucleation centers, resulting in a direct correlation with the size of the crystallites in Fig. S1C. The room temperature XRD patterns of Fig. S1B confirm not only the phase purity of these samples, but also the excellent crystallinity.

Crystals larger than 20 microns are obtained when the reaction between $CuSeO_3 \cdot 2H_2O$ and $NH_4OH$ proceeds in an autoclave at 150 °C (Fig. 2B). The "Bulk" sample was obtained by crushing few single crystals of $Cu_2OSeO_3$, similar to the one seen in Fig. 2A which is grown by chemical vapor transport.[35]

All the in-house XRD measurements were carried out on a PANalytical Empyrean diffractometer with the Cu-K$_\alpha$ radiation. The *in situ* XRD measurements were carried out at the Swiss-Norwegian Beamlines (SNBL) at the European Synchrotron Radiation Facility (ESRF), Grenoble, France. The X-ray wavelength used for these experiments was $\lambda = 0.69$ Å. HRTEM investigations were conducted using Thermo Fisher Scientific Talos TEM at 200kV accelerating voltage. X-ray fluorescence (XRF) measurements were performed on an ORBIS PC with the applied voltage of 26 kV and with a current of 1000 $\mu$A. The quantification was done using the K$\alpha$ line of both Cu and Se. For a good statistical distribution, measurements were performed at 128 different points for each sample. Raman scattering measurements were performed on a Renishaw inVia Raman spectroscope equipped with a 532 nm wavelength laser. The incident power on the sample was kept fixed at 0.05 mW (spectral resolution of 0.8 cm$^{-1}$). Pellets were made from each sample and ten repetitions were performed (at different positions), each with a 100 seconds of exposure time.

**Magnetic measurements.** All the DC as well as AC magnetic measurements were performed using the VSM and ACMS-II option of a commercial Quantum Design (QD) 14T Physical Property Measurement System (PPMS). The powder sample was enclosed in the standard polypropylene holder provided by QD. For the AC measurements, the excitation field and frequency were set to be at 0.1 mT and 1000 Hz respectively (unless explicitly specified). To reduce the hysteretic effects, samples were demagnetized in the paramagnetic regime (70 K) after each magnetic field scan.

**Small Angle Neutron Scattering.** The long period magnetic structure in the Cu$_2$OSeO$_3$ particles of two sizes, 338 nm and 73 nm, were measured using SANS-II instrument at the Swiss Spallation Neutron Source (SINQ), Paul Scherrer Institute (PSI), Switzerland and D33 instrument at the Institut Laue Langevin (ILL), Grenoble, France.

Samples for experiments at both instruments were prepared as pressed pellets of 10 mm diameter and thickness <0.5 mm and installed into a horizontal field cryomagnet. The magnetic field was applied in the plane of the sample pellet, and perpendicular to the neutron beam. For the field training procedure, the sample was first saturated to field-polarized state while collecting statistics at specific magnetic field values shown in the main text. This was followed by ramping down the magnetic field to 0 mT at a rate of 10 mT per second.

At SANS-II, neutrons with a wavelength of 7.29 Å were used with a FWHM spread $\Delta\lambda/\lambda$=10 %. The incoming beam was collimated over a distance of 6 m before the sample, with the scattered neutrons collected by a two-dimensional multidetector placed 6 m behind the sample. At D33, neutrons with a wavelength of 4.6 Å were used with a FWHM spread $\Delta\lambda/\lambda = 10\%$. Both the collimation distance and the sample-to-detector distance were 12.8 m. The data collected at D33 are available from the ILL.[36] All the data analysis was performed using the GRASP software.[37]

**Micromagnetic simulations.** Landau-Lifshitz-Gilbert (LLG) simulations calculations were performed at $T = 0$ K using the MuMax3 software package, based on finite difference discretization,[38] and appropriate magnetic parameters for a Cu$_2$OSeO$_3$ particle: exchange stiffness $A_{ex}$ = 7e-13 Jm$^{-1}$ and Dzyaloshinskii-Moriya interaction constant D = 7.4e-5 Jm$^{-2}$.[39,40] The size of the elementary unit of 4 nm was used. An initially random spin configuration was allowed to relax for 2.5 ns in each simulation. The magnetic field was first applied along the top vertex of the octahedron shown in Fig. 7A. Similar calculation was performed for when the external magnetic field was applied perpendicular to one of the octahedron facets, results of which is shown in Fig.

7B. Note that the various spin textures are identified by observation only, since the definition of topological charge breaks down in 3D.[41] The computation was performed using NVIDIA GPU 980Ti, and RTX 3080 Ti. The magnetic structure was visualized using the Paraview software.[42]

## RESULTS AND DISCUSSION

**Chemical synthesis and structural analysis of $Cu_2OSeO_3$ particles.** Adapting the concept of selenious acid titration to selenite, $Cu_2OSeO_3$ particles were synthesized with an excellent control over the stoichiometry, crystal structure and size (Fig. 1). Exposing a copper selenite precursor to a basic aqueous solution induces its chemical decomposition through the leaching of highly water soluble selenious acid. The amount of leached $H_2SeO_3$, *i.e.* the final Se stoichiometry of the selenite, can be controlled by the available amount of base to be neutralized. $CuSeO_3 \cdot 2H_2O$ used as precursor is then fully converted into $Cu_2OSeO_3$ when the stoichiometry of the equation 1 is realized:

$$2\ CuSeO_3 \cdot 2H_2O + 2OH^- \rightarrow Cu_2OSeO_3 + SeO_3^{2-} + 5H_2O \qquad \text{(eq. 1)}$$

The absence of an excess of basic species in the reaction medium prevents $Cu_2OSeO_3$ decomposing further into CuO. The reaction proceeds as long as all the available basic species are neutralized and pure $Cu_2OSeO_3$ particles are obtained, see Fig. 1B and Supplementary Fig. S1A. Full details of the particle growth are found in Methods. To ensure the controlled growth of the nucleation centers, which correlates directly with the resulting particle size, relatively low temperatures (70 °C – 100 °C) were used for the reflux process of the conversion reaction. Using this in-solution growth, samples of $Cu_2OSeO_3$ particle with eight different sizes were synthesized.

The crystal structures and chemical compositions of the various particle samples can be compared with those of a standard $Cu_2OSeO_3$ powder obtained from a ground bulk single crystal grown by the often-used technique of chemical vapor transport, see Fig. 2A. As shown in Supplementary Fig. S1, powder x-ray diffraction (XRD) reveals the synthesized $Cu_2OSeO_3$ samples to display excellent purity and crystallinity. This is also corroborated by the selected area electron diffraction images shown in Fig. S5. The crystallization of the particles is inferred from the peak broadening in the XRD patterns data (see Supplementary Fig. S1C) and from scanning electron microscopy images, see Fig. 2C-D, which reveal all particles to display an octahedral morphology. Fig. 2E shows a Williamson Hall (W-H) analysis of the XRD data, from which the average coherent particle size is determined and the nanoparticle strain is estimated. The various samples of $Cu_2OSeO_3$ particles have octahedral heights ($h$) ranging from 73 nm to 338 nm and negligible strain (< 0.2%). Further details about particle size determination is described in the supplementary material section, including Figs. S2 and S3. Raman spectroscopy data shown in Fig. 2F confirm the structural equivalence between the differently sized particles and the ground bulk sample. From X-Ray Fluorescence (XRF) measurements, the Cu and Se stoichiometry of the particles was also investigated. Fig. 2G shows that average Cu/Se atomic ratio is determined to be 1.99(2) and 2.01(1) for the 73 nm and 338 nm particles, respectively. As shown in Fig. S6, by using high resolution transmission electron microscopy (HRTEM), the particles were found to be highly crystalline and free of defects. Additional refinement of the powder XRD patterns, shown in Fig. S4, confirms that the unit cell length for all particle sizes remain essentially unchanged from the bulk value. In summary, volume aside, the synthesized particles are essentially chemically and structurally equivalent to bulk $Cu_2OSeO_3$.

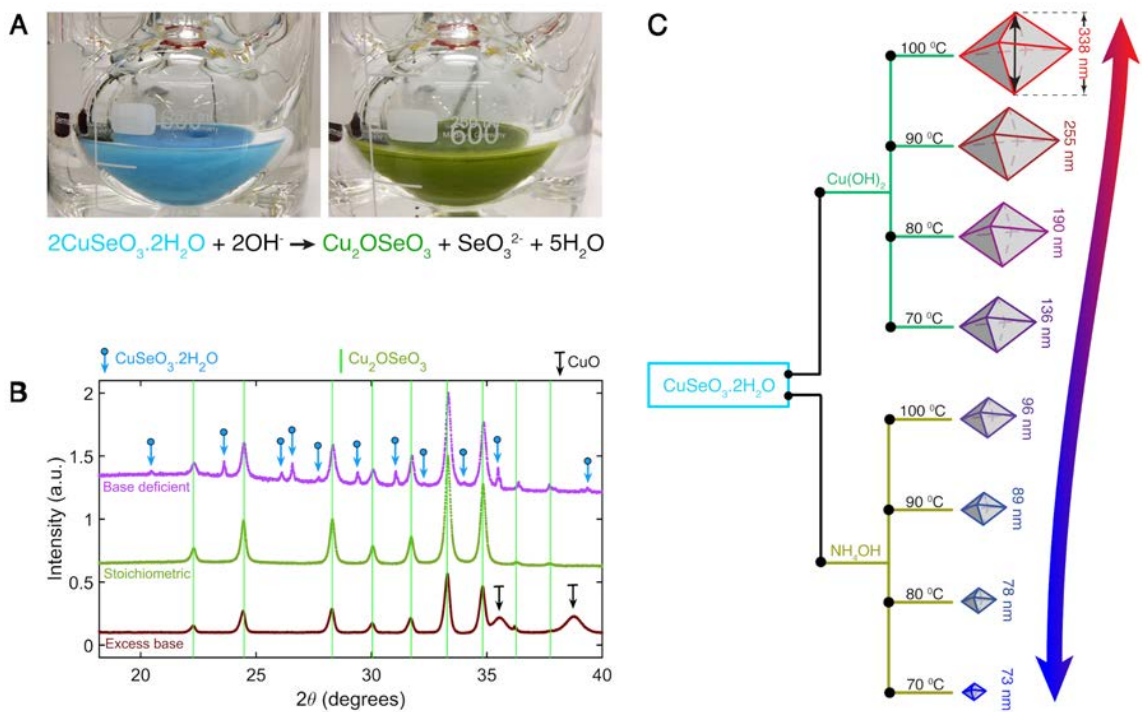

**Figure 1.** Overview of $Cu_2OSeO_3$ synthesis. (A) Optical images of the starting solution containing $CuSeO_3 \cdot 2H_2O$ on the left, and the final product, $Cu_2OSeO_3$ on the right. In a 250 ml round bottom flask, about 1.5 g of $Cu_2OSeO_3$ particles are produced in about 5 hours at 100 °C. The chemical equation describes the solid-to-solid transformation between the initial and final products in presence of a basic medium (see Fig. S1A). (B) X-ray diffractograms of the final products as obtained from the reaction when stoichiometric and off-stoichiometric mixture of $CuSeO_3 \cdot 2H_2O$ and base were used. Using an excess of base results in the presence of CuO formed by the decomposition of $Cu_2OSeO_3$ ($Cu_2OSeO_3 + 2OH^- \rightarrow 2CuO + SeO_3^{2-} + H_2O$). When the base is deficient, part of $CuSeO_3 \cdot 2H_2O$ remains unreacted. (C) Schematic illustration linking synthesis protocol and the final average particle size as obtained from X-ray analysis (see Fig. 2E).

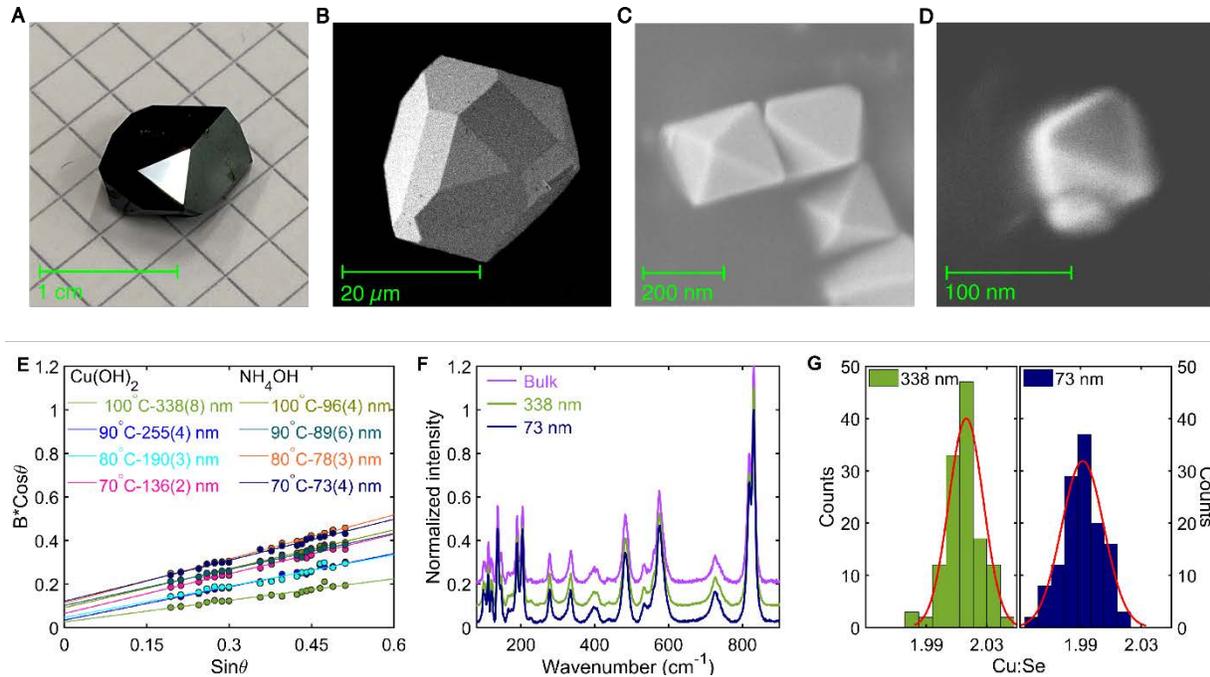

**Figure 2.** Chemical and structural characterizations of Cu$_2$OSeO$_3$ particles. (A)-(D) Representative images of the particles discussed in the text. (E) Williamson-Hall analysis for all the particles. The instrumental contribution of the peak broadening was determined in the XRD pattern recorded with the bulk Cu$_2$OSeO$_3$ powder. The slope and intercept of the fits are representative of the average particle size and strain, respectively. The different particle sizes obtained are given with the corresponding reflux temperatures and bases used during the synthesis. (F) Raman spectrum of selected particles showing distinct modes consistent with the ones obtained from the bulk sample. These results are representative of those from remaining sizes of particles. (G) XRF analysis of the same selected Cu$_2$OSeO$_3$ particles showing the atomic ratio of Cu and Se being very close to the expected stoichiometric value of 2.

**Magnetic phase diagrams of Cu$_2$OSeO$_3$ particles.** The magnetism of Cu$_2$OSeO$_3$ is governed by S=1/2 Cu$^{2+}$ ions that are oxygen anion coordinated in square pyramids or trigonal bipyramids. Due to the chiral cubic lattice symmetry, the antisymmetric Dzyaloshinskii-Moriya interaction (DMI) is also active, and in combination with the symmetric exchanges, plays the key role leading to the formation of both the helical ground state and skyrmion phases. Each is characterized by a modulation period of 60 nm in the bulk,[14,43-44] which is much larger than the cubic crystal lattice constant of 0.89 nm. Our samples display length-scales between slightly greater than one bulk helical pitch (73 nm) up to larger than five (338 nm) (Fig. 2). Fig. 3 summarizes isothermal AC susceptibility data obtained on a bulk sample, along with particles with octahedral heights of 338 nm, and 73 nm. Data for other particle sizes are presented in Supplementary Fig. S7. At 57 K in the bulk sample, shown in Fig. 3B, the standard sequence of field-driven phase transitions from helical (Hel.) → conical (Con.) → skyrmion lattice (SkL) → Con. → field-polarized (FP) are observed in the data. A characteristic two-peak structure in the imaginary ($\chi''$) part of the AC susceptibility is seen clearly between 18 and 30 mT, signaling the range of SkL phase stability in analogy with other chiral magnets.[45-48] The established bulk phase diagram, see Fig. 3C, shows that the SkL phase extends over a narrow temperature window of ~3 K directly below $T_c$.

By tracking the two-peak feature in $\chi''$ for in 338 nm particles (Fig. 3E), the thermal window of SkL phase stability is found to extend to 10 K below $T_c$, as shown in Fig. 3F. In the established absence of particle strain, the enhanced thermal stability is likely due to the onset of geometrical confinement effects analogous to those observed in thin layers of cubic chiral magnets.[13,49] As additional evidence for confinement effects, further magnetometry data presented in Supplementary Fig. S8 reveals the systematic suppression of $T_c$ as $h$ is reduced. In bulk $Cu_2OSeO_3$, it is established that pronounced paramagnetic spin fluctuations already suppress the onset of helical order at $T_c$ to a slightly lower temperature than anticipated for a conventional ferromagnet.[50] The further reduction of $T_c$ with $h$ thus indicates geometrical confinement to effectively suppress the density of fluctuations in the system. A similar behavior is also seen in FeGe nanoparticles,[27] which suggests the phenomenon to be a generic feature of chiral magnets as their size approaches the nanoscale.

In the $h = 73$ nm particles, no signatures in AC susceptibility delineating a skyrmion phase are observed. Instead, a differently structured phase diagram is found with a lower $T_c$ and a single low field phase boundary giving a clear signal in $\chi''$ (Fig. 3H and I). The dissipative nature of the phase boundary may indicate a transformation between imperfect, large-scale spin configurations, which the simulations presented later suggest are unlikely to be due to skyrmions. According to the established phase diagrams for $h$ ranging from 338 nm to 73 nm (Supplementary Fig. S7), the threshold octahedral height ($S_c$) for which standard signatures of the skyrmion are observed is 190 nm. For smaller sizes below $S_c$, the phase diagrams resemble that shown in Fig. 3I, hosting helical-like, but unestablished spin configurations labelled Phase-I and Phase-II.

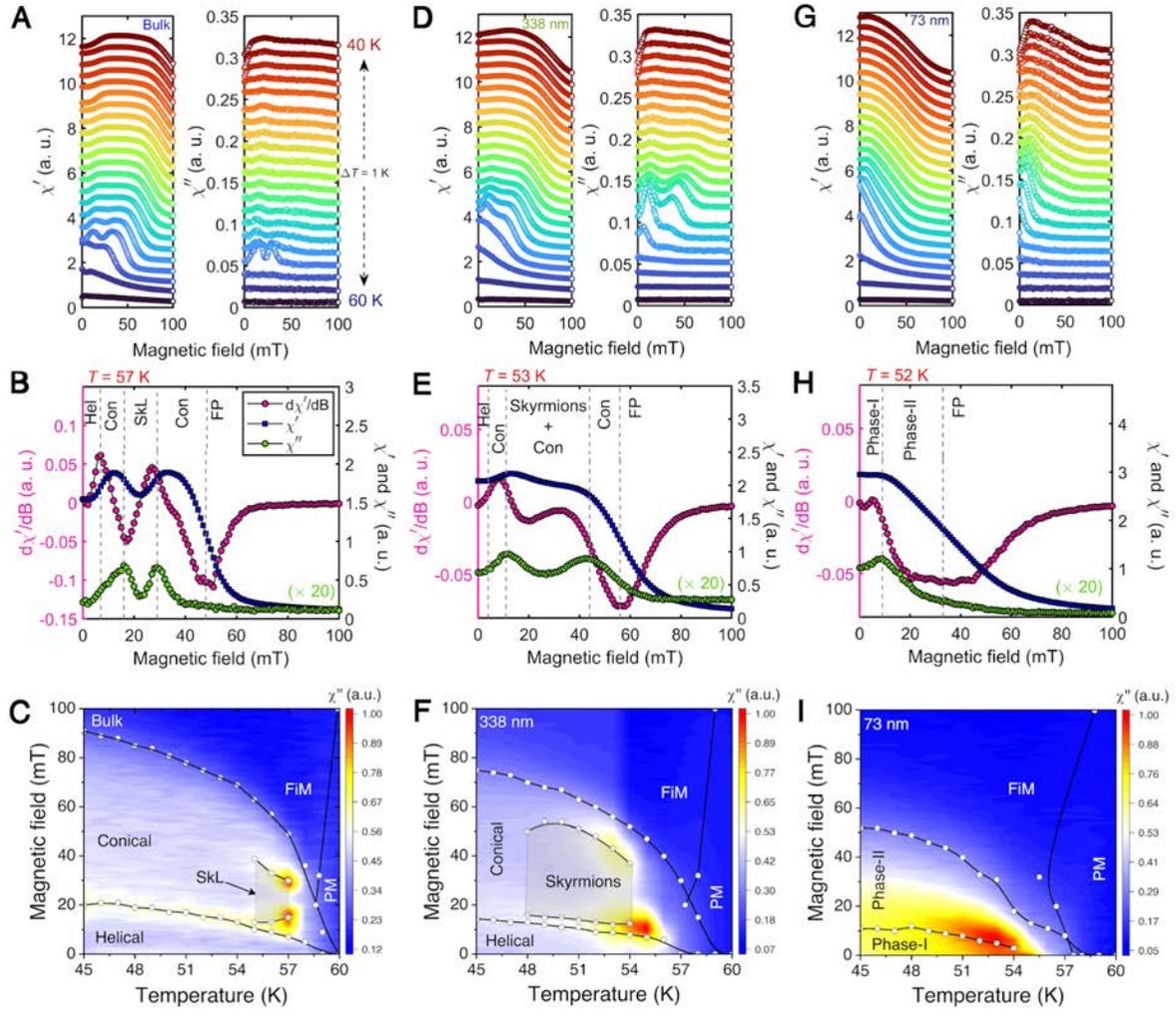

**Figure 3.** AC susceptibility of Cu$_2$OSeO$_3$ particles. (A), (D) and (G) show AC susceptibility field scans for three different particle sizes between 40 K and 60 K, with 1 K step. Clear differences can be observed between these particles from the imaginary part of the susceptibility, especially the extended "valley" region in the 338 nm particles at 53 K. The same feature is entirely absent in the smallest particles, 73 nm. (B), (E), and (H) show the procedure of extracting the boundaries between different phases from both the real part χ´ and imaginary part χ˝ of the susceptibility. Note that χ˝ data has been scaled up in order to show all in the same graph. (C), (F), (I) represent the magnetic phase diagrams thus extracted from the χ˝ data for the respective particle sizes.

**SANS observation of skyrmions and confined magnetic modulations.** Small angle neutron scattering (SANS) experiments were performed to elucidate the magnetic textures in $Cu_2OSeO_3$ particles with $h$ of 338 nm and 73 nm, bigger and smaller than $S_c$, respectively. Full details of the SANS experiments are given in the Methods, with experimental setup illustrated in Fig. 4A. Fig. 4B summarizes the connection between the real-space magnetic textures (left column) consistent with those found in bulk chiral magnets, and the corresponding SANS patterns (center and right columns) when such textures are stable in particles that are unoriented with respect to one another. In the 338 nm particles at 52.4 K and after zero field cooling, Fig. 5A shows the appearance of a ring-like magnetic scattering around $Q = 0$ due to propagating helical spin correlations within a sample of the randomly oriented particles. From the intensity ring, the magnitude of the wavevector describing the magnetic modulation is determined as $Q = 0.1027(2)$ $nm^{-1}$ (Supplementary Fig. S9). This gives a magnetic modulation periodicity ($\lambda$) of 61.2(1) nm consistent with that due to helical order found in bulk samples,[20,43,44] as illustrated in Fig. 4B.

Fig. 5B-F show that a horizontal magnetic field applied perpendicular to the incident neutron beam drives a rearrangement of the SANS intensity in the 338 nm ($> S_c$) particles. As indicated in Fig. 4B, a redistribution of the SANS intensity to regions aligned with the field (green boxes in Fig. 5A-F) and perpendicular to the field (red boxes in Fig. 5A-F) are hallmark signatures for the formation of conical and skyrmion orders, respectively.[53] The formation of both of these orders is most clearly seen in Fig. 5C, where we observe peaks due to co-existing conical and skyrmion orders consistent with those expected according to Fig. 4B. Fig. 5G shows the field-dependence of the green box (helical and conical modulations) intensity to display a similar field-dependence and two-peak structure as that of $\chi''$ determined from AC susceptibility. These are further clear signs for skyrmion formation in the "valley" field region $15 < B < 35$ mT. Concomitantly, over the same field region, Fig. 5H reveals a small, yet clear enhancement of the red box (skyrmion) SANS intensity above any residual ring intensity (white boxes in Fig. 5A-F) that remains undistributed by the field before saturation, presumably due to pinning. In the middle of the valley region shown in Fig. 5G, the intensity of the green boxes drops by roughly 35%, while a similar percentage increase of intensity in the red boxes compared with the white boxes is also observed in Fig. 5H. These results indicate that ~35% of the particles in the sample undergo a transition to a skyrmion-like phase. and that these coexist with particles that remain in unoriented helical-like and conical like phases. Overall, these SANS data confirm microscopically the existence in the 338 nm particles of phases directly analogous to those found standardly in bulk chiral magnets.

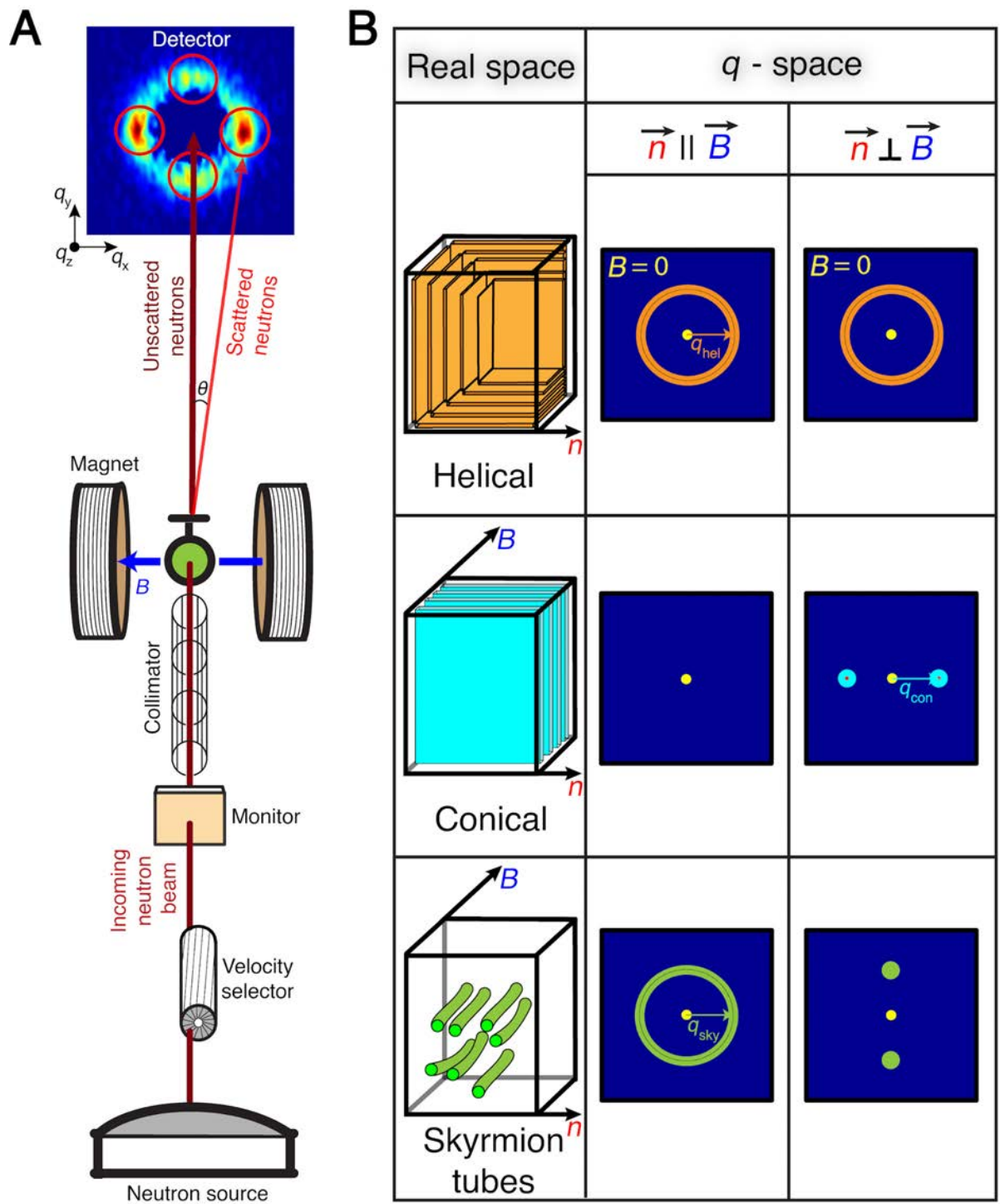

**Figure 4.** Proposed magnetic textures in real and reciprocal space. (A) Schematic illustration of the experimental geometry used in our SANS measurements. The applied magnetic field was always kept perpendicular to the incoming neutron beam. The detector image shown is the scattering pattern obtained for the 338 nm particles, in an external field of 20 mT and a temperature of 52.4 K. (B) Left column: Proposed chiral magnetic textures that may be stable inside each 338 nm particle, such as multidomain helical phase, single domain conical phase and skyrmion phase. Their corresponding reciprocal space

scattering patterns for a magnetic field aligned with the neutron beam (center column) and perpendicular to the neutron beam (right column) are given when such textures are stable inside a sample of unoriented particles. The central yellow spot is due to the unscattered neutrons. These have been masked in the real detector images (such shown as in A or Fig.5). The helical phase stable in zero field shows up as a sphere of intensity in reciprocal space, and correspondingly always as a ring on the detector. In the conical phase, the magnetic order always propagates along the magnetic field, and so the scattering is concentrated into peaks that are observed only when the neutron beam is perpendicular to the field. Since skyrmion tubes align with the field direction, the scattering signal due to skyrmions in the sample of unoriented particles, manifests as a ring in the plane perpendicular to the applied field, and thus presents as two peaks at the north and south pole positions when the field is perpendicular to the neutron beam.

In the 73 nm ($< S_c$) nanoparticles at 52.4 K near to $T_c$, a ring of magnetic SANS intensity is again observed around $Q = 0$ after zero-field cooling (Supplementary Fig. S10) from which the pitch length λ, was determined to be 72(1) nm, as shown in Supplementary Fig. S10I. The value of λ is longer than that of ~60 nm expected for helical order in bulk samples, and spans exactly the average octahedral height of the nanoparticle. The adaptation of the ground state magnetic modulation in accordance with the geometry of the nanoparticles is a clear sign for a finite size effect. Furthermore, as discussed in the Supplementary Material and shown in Supplementary Fig. S11, a ring-like scattering pattern remains consistent with that expected from a sample of unoriented particles each hosting a single-period helical spiral structure that spans the octahedral particle height. Under increasing magnetic field, the confinement effect on λ is relieved as it falls towards the bulk value for helical order on the approach to saturation, shown in Supplementary Fig. S10I. In addition, there is no observed rearrangement of the intensity in the ring, and instead the SANS intensity falls monotonically on the approach to saturation, see Supplementary Fig. S10H. This latter observation may be indicative of either a pinning effect on the magnetic modulations, or the realization of an unexpected spin configuration with more isotropic susceptibility that is without analogy to the standard chiral magnet phases realized in the bulk.

Finite size-effects on λ are also observed far from $T_c$ at 5 K in both the 73 nm and the 338 nm particle samples. Fig. 5I and J each show two-spot SANS patterns indicative of helical-like spin-correlations existing in each sample at zero field after applying a standard field-training procedure performed at 5 K. As shown in Fig. 5 K, the magnetic periodicity determined from the spot positions is λ = 72(2) nm in the 73 nm nanoparticles, which is the same as close to $T_c$ (Supplementary Fig. S10I). For the 338 nm particles, λ = 65(1) nm is larger than close to $T_c$, and has expanded so that exactly five magnetic modulations span the octahedral height. We note that the observed enhancement of λ in the particles at low temperature is an order of magnitude larger than the increase otherwise observed in bulk $Cu_2OSeO_3$.[54]

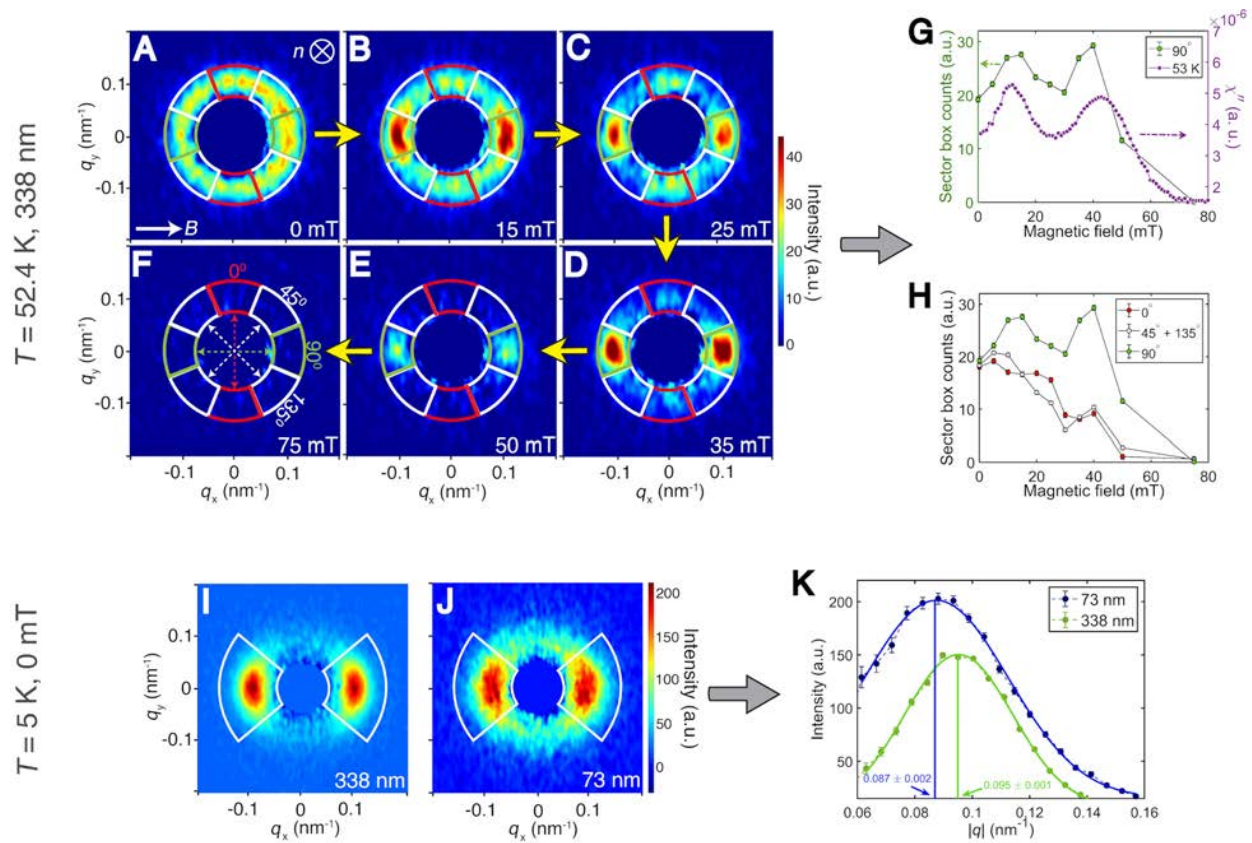

**Figure 5.** Probing $Cu_2OSeO_3$ particles using SANS. (A)-(F) SANS detector images from the $h$ = 338 nm particles measured at 52.4 K and as a function of the applied magnetic field. For simplicity, the intensity scale has been kept the same for all patterns. The angular positions of sector boxes indicated in panel F represent sector boxes corresponding to that particular position, as well as the one which is situated diametrically opposite to it. (G) shows the evolution of intensity in the sector boxes corresponding to the conical phase [green boxes in (A)-(F)], as a function of applied magnetic field. For a comparison, $\chi''$ for the same particles measured at 53 K is also shown. (H) The dip in conical phase SANS intensity between 15 and 35 mT indicates a redistribution of the intensity due to the formation of skyrmion tubes. (I) and (J) show the detector images obtained in zero field and 5 K after performing a field training procedure for both particle sizes (Method section). Here the zero-field state was prepared by ramping the magnetic field down from saturation in an attempt to concentrate the SANS intensity into a smaller reciprocal space volume. This hysteric effect of field training on the spiral propagation vector is typical for cubic chiral magnets at low temperatures.[51,52] The wavevector-dependence of the resulting intensity indicated by the sectors in (I) and (J) is shown in (K), and the peak positions determined by solid lines.

**Particle size-dependent magnetic configurations.** To understand better the role of the particle geometry and size on the stability of different magnetic configurations, micromagnetic simulations were performed for a realistic octahedron-shaped particle, and exchange, DMI and magnetization constants typical for bulk $Cu_2OSeO_3$ (see Methods).[39] The resulting variety of magnetic textures found by simulations of a single particle are expected to be representative for our samples, since the dipolar interaction is shown to be negligible for these systems, as pointed out recently by means of finite-element simulations.[55] In the absence of an external magnetic field, the initially random spin configuration relaxes to a state with randomly-oriented helices for all particle sizes, as shown in Fig. 6A, E, and I. Under an external magnetic field applied along the octahedron height, the magnetization of the smaller octahedra ($h$ = 100 - 220 nm) gradually rotates towards the field direction without developing any non-trivial magnetization textures (Fig. 6A-D). A similar smooth transformation from a zero-field non-collinear to a field-induced ferrimagnetic state was reported previously in a study of very small ($h \sim 30$ nm) $Cu_2OSeO_3$ nanoparticles.[26]

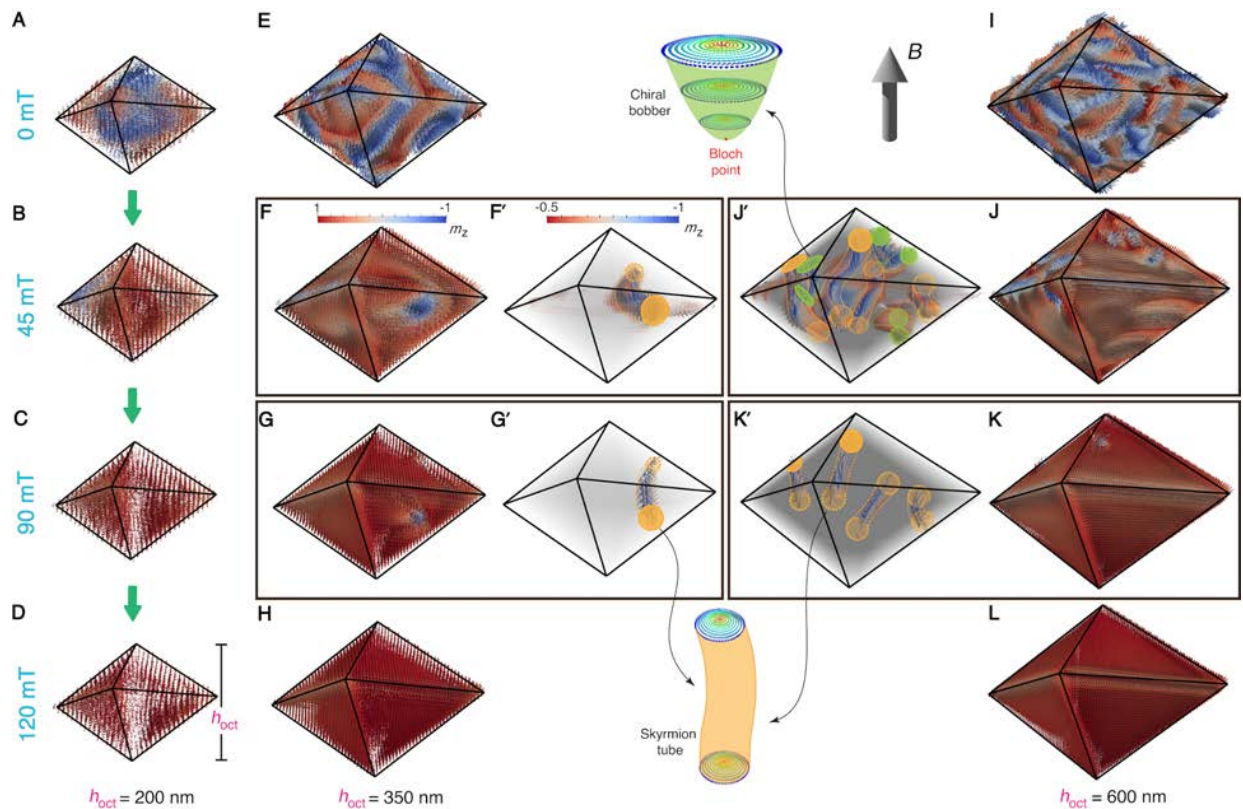

**Figure 6.** Micromagnetic simulations on isolated octahedral $Cu_2OSeO_3$ particles. (A)-(D), (E)-(H) and (I)-(L) show magnetic textures obtained from the micromagnetic simulation performed on octahedrons of height 200 nm, 350 nm and 600 nm, respectively. Panels (F´, G´) and (J´, K´) elucidate the internal magnetic structures of 350 nm and 600 nm particles in magnetic fields of 45 and 90 mT, respectively. The black lines have been added in order to make the particle contours more obvious. The appearance of chiral bobbers and skyrmion tubes is highlighted by green and yellow ellipses, respectively. The direction of applied magnetic field was kept the same in all cases, which is along the top vertex of the octahedron.

In contrast, larger octahedra exhibit a variety of field-induced magnetic textures, including half-skyrmions (merons) in octahedra with 220 < $h$ < 350 nm, chiral bobbers (350 < $h$ < 400 nm) and skyrmion tubes ($h$ > 400 nm) in various (field-dependent) co-existent combinations. Representative simulations for particles with $h$ = 350 nm and $h$ = 600 nm are shown in Figs. 6E-L, wherein skyrmion and chiral bobber spin textures are stabilized. The field-induced transformation of the helical structure to the meron state in the particles allows a minimization of the Zeeman energy while preserving the periodicity of the magnetic texture, when formation of the complete skyrmion tube is unfavored by the confinement effect. The chiral bobber state, seen as a combination of a skyrmion tube and a Bloch point, is stabilized for the same critical particle size as the skyrmion state, suggesting similar formation mechanisms for these two textures. Our simulations suggest stabilization of chiral bobber state as an edge state of $Cu_2OSeO_3$ in larger nanoparticles which is consistent with generalized theory of chiral bobbers at the surface in cubic chiral magnets and experimental works on FeGe which hosts helimagnetic ground state with a spiral pitch similar to $Cu_2OSeO_3$.[56-58] Recently, the bobber phase induced by proximity effect was also evidenced in $Cu_2OSeO_3$ / magnetic multilayer heterostructure.[59] However, their unambiguous experimental evidence in bulk and nanostructured $Cu_2OSeO_3$ is remaining as a challenging task which can be further addressed by real-space electron holography or resonant elastic x-ray scattering. Further, we find that in $Cu_2OSeO_3$ nanocrystals the critical size for nucleation of a single skyrmion tube was found to be around 350 nm (Fig. 6F and G). The phase diagram summarizing these findings is shown in Fig. 7A.

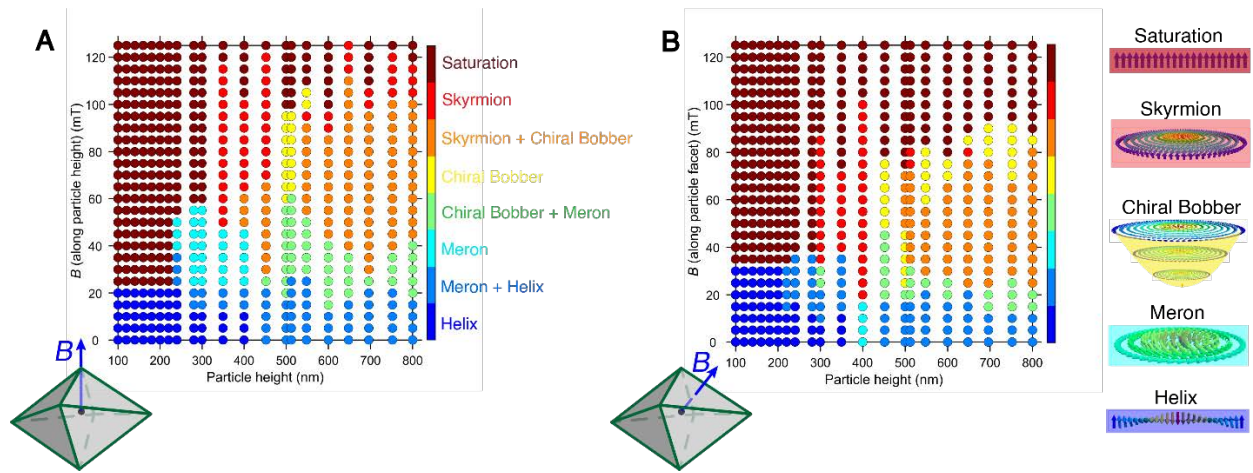

**Figure 7.** Size dependent phase diagram of octahedral $Cu_2OSeO_3$ particles. Spin texture phase diagram as a function of height of the octahedron, as obtained from micromagnetic simulations. (A) The magnetic field was applied along the top vertex of the octahedron, that is along [001] whereas, in (B) the magnetic field was applied along one of the triangular facets, that is along [111]. The dotted lines represent the simulation points in the phase diagram. All the magnetic textures were identified manually.

Similar simulations were done for the case when the magnetic field was oriented perpendicular to one of the octahedral triangular facets. The size-field phase diagram (Fig. 7B) appears qualitatively similar to Fig. 7A, with small differences in bobber and skyrmion stability regions. For each orientation of the magnetic field, a similar threshold volume of magnetic particle is identified below which topologically non-trivial magnetization configurations analogous to those found in the bulk are not expected to be stable. The field evolution of the magnetic texture is qualitatively reflected in the magnetization curves directly extracted from the micromagnetic simulations (Supplementary Fig. S12). These findings agree with the experimental magnetometry and SANS data, and the expectation for other particle shapes considered in previous micromagnetic simulations.[26,28-29,55,60-61] On the other hand, the present experimental data do not allow a straightforward identification of the different topological spin textures expected according to our simulations. Real-space imaging probes such as electron holography or Lorentz-transmission electron microscopy (LTEM) are needed to determine the detailed spin configurations at the single particle level (see supplementary Fig. S13).

**CONCLUSION**

Size-tailored octahedral $Cu_2OSeO_3$ particles approaching nanoscale sizes were synthesized by a low temperature process, and shown to be chemically and structurally equivalent to bulk $Cu_2OSeO_3$. The chemical treatment is based on selenious acid leaching from a selenite precursor induced by an acid-base neutralization. We anticipate that the same chemistry can be used to grow other selenite materials. Likewise, it can be applied to selenates when exposed to a base by leaching of selenic acid. Selenates and selenites are an important class of materials with a large spectrum of applications involving optical properties, catalysis, energy to name but a few. From magnetometry and neutron scattering experiments, together with micromagnetic simulations, we explored the interplay between the particle geometry and chiral magnetic interactions in $Cu_2OSeO_3$, and find that a diverse array of magnetic configurations with variable topology may be stabilized according to the particle size. Correspondingly, our results shed new and detailed light on the size-dependence of the magnetic phase diagrams and magnetization textures in small $Cu_2OSeO_3$ samples. This includes the discovery for the first time of a critical octahedral height $S_c$ of 190 nm ($3\lambda$) below which no standard experimental signature for skyrmion tubes is found. For other types of topological spin textures (e.g. meron, bobber), expected to be stable for shorter octahedral heights, investigation using experimental techniques such as LTEM and electron holography is highly desired. Elucidation of the magnetic textures in the larger particles also calls for the further development of neutron and x-ray beam techniques aiming at the reconstruction of the real-space magnetization distribution inside the particle.[62,63]

The observed quantization effect of confined chiral magnetic modulations was revealed previously in epitaxially- grown $MnSi$[64,65] and $FeGe$[66,67] chiral magnet thin films. In these systems, the periodicity of helical-like phases was found to either increase or decrease slightly with respect to the bulk so that a discrete integer or half-integer number of helical turns span between film surfaces. The present work extends observation of this phenomenon to small $Cu_2OSeO_3$ particles, where for each of $h = 73$ nm and 338 nm we observe clear increases in $\lambda$ so that a single and five helical turns span the octahedron, respectively. The discretization in $\lambda$ is generally explained as enforced by boundary conditions due to the reduced dimensionality and the surface-induced magnetic anisotropy. In contrast to spanning flat surfaces, the confined low temperature integer-turn helical-like states span the octahedral height, or more generally, the non-adjacent octahedral vertices where four octahedron edges meet. This shows the boundary conditions enforced by the geometry of a vertex, itself a form of structural topological defect, become increasingly dominant in the limit of small particle volume, low temperature and low field. Moreover, since in the vicinity of the particle vertices, edges and surfaces, the symmetry of the magnetic interactions is naturally broken compared with the bulk, entirely new magnetic configurations without analogy to the bulk have the potential to emerge. Due to this key role of the geometry in generating the interactions that underlie the confined magnetic configurations, their robust stability can be anticipated, particularly in the genuine nanoparticle limit. The geometrical confinement effects reported here bring fresh insight for the future exploitation of chiral magnets with sizes approaching the nanoscale, in particular by defining relevant length-scales for their use in applications settings.

## SUPPORTING INFORMATION

High resolution synchrotron XRD patterns of *in situ* transformation between the precursor $CuSeO_3 \cdot 2H_2O$ and final product $Cu_2OSeO_3$, High-resolution room temperature XRD patterns of all the particles, DLS measurements, SAED image, HRTEM analysis, magnetic phase diagrams of all particle sizes, Zero-field AC susceptibility temperature scans for all the particles, SANS detector image and magnetic scattering as a function of *q* for 338 nm particles at 53 K at zero magnetic field, SANS detector images as well as magnetic scattering as a function of magnetic field for the smallest nanoparticles (73 nm), illustration of a single period of magnetic helix embedded into 73 nm particle as well as a Fast-Fourier transform of the *z*-component of the magnetization, total magnetization as a function of magnetic field, as obtained from micromagnetic simulations (PDF). The dataset and code used for this article are available on zenodo.org.[68]

## AUTHOR CONTRIBUTIONS

A.M. and J.S.W. conceived the project. P.R.B., J.S.W. and A.M. designed the experiments. P.R.B. and A.M. synthesized the nanoparticles and performed the structural and chemical characterizations. P.R.B. and I. Ž. performed the magnetometry measurements. SEM and TEM measurements were performed by P.R.B. & T.L., and analyzed by T.L. The SANS measurements were carried out by P.R.B., V.U., R.C., H.M.R., and J.S.W. V.U. and P.R.B. performed the micromagnetic simulations which were analyzed by V.U. All experimental data, except HRTEM, were analyzed by P.R.B. P.R.B., V.U., I. Ž., H.M.R., J.S.W. and A.M. discussed the data. P.R.B., V.U., J.S.W. and A.M. wrote the manuscript. All authors commented on the manuscript.

## ACKNOWLEDGMENT

Funding from the Swiss National Science Foundation (SNSF) Sinergia network NanoSkyrmionics (Grant No. CRSII5_171003) and Project No. 200021_188707 are gratefully acknowledged. We thank Kiyou Shibata and Oleg V. Yazyev for their support and helpful discussions. This work is based on experiments performed at the Swiss spallation neutron source (SINQ), located at the Paul Scherrer Institute, Villigen, Switzerland, the Institut Laue-Langevin (ILL), Grenoble, France, and the Swiss-Norwegian Beamline (SNBL) at European Synchrotron Radiation Facility (ESRF), Grenoble, France.

**REFERENCES**

(1) Geim, A. K.; Novoselov, K. S. The Rise of Graphene. *Nat. Mater.* **2007**, *6*, 183-191.

(2) Bedanta, S.; Barman, A.; Kleemann, W.; Petracic, O.; Seki, T. Magnetic Nanoparticles: A Subject for Both Fundamental Research and Applications. *J. Nanomater.* **2013**, 952540.

(3) Matatagui, D.; Kolokoltsev, O. V.; Qureshi, N.; Mejía-Uriarte, E. V.; Saniger, J. M. A Magnonic Gas Sensor Based on Magnetic Nanoparticles. *Nanoscale* **2015**, *7*, 9607-9613.

(4) Kosterlitz, J. M.; Thouless, D. J. Ordering, Metastability and Phase Transitions in Two-Dimensional Systems. *J. Phys. C Solid State Phys.* **1973**, *6*, 1181-1203.

(5) Parkin, S. S. P.; Hayashi, M.; Thomas, L. Magnetic Domain-Wall Racetrack Memory. *Science* **2008**, *320*, 190–194.

(6) Chemistry and physics of nanowires; Xia, Y., Yang, P., Eds.; *Wiley Online Library*, 2003.

(7) Kodama, R. Magnetic Nanoparticles, *J. Magn. Magn. Mater.* **1999**, *200*, 359–372.

(8) Batlle, X.; Labarta, A. l. Finite-Size Effects in Fine Particles: Magnetic and Transport Properties. *J. Phys. D* **2002**, *35*, R15–R42.

(9) Fernández-Pacheco, A.; Streubel, R.; Fruchart, O.; Hertel, R.; Fischer, P.; Cowburn, R. P. Three-Dimensional Nanomagnetism. *Nat. Commun.* **2017**, *8*, 1–14.

(10) Iwasaki, J.; Mochizuki, M.; Nagaosa, N. Current-Induced Skyrmion Dynamics in Constricted Geometries. *Nat. Nanotechnol.* **2013**, *8*, 742–747.

(11) Wang, X. S.; Yuan, H. Y.; Wang, X. R. A Theory on Skyrmion Size. *Commun. Phys.* **2018**, *1*, 1–7.

(12) Mühlbauer, S.; Binz, B.; Jonietz, F.; Pfleiderer, C.; Rosch, A.; Neubauer, A.; Georgii, R.; Böni, P. Skyrmion Lattice in a Chiral Magnet. *Science* **2009**, *323*, 915–919.

(13) Yu, X. Z.; Kanazawa, N.; Onose, Y.; Kimoto, K.; Zhang, W.Z.; Ishiwata, S.; Matsui, Y.; Tokura, Y. Near Room-Temperature Formation of a Skyrmion Crystal in Thin-Films of the Helimagnet FeGe. *Nat. Mater.* **2011**, *10*, 106–109.

(14) Seki, S.; Yu, X.Z.; Ishiwata, S.; Tokura, Y. Observation of Skyrmions in a Multiferroic Material. *Science* **2012**, *336*, 198–201.

(15) Kézsmárki, I.; Bordács, S.; Milde, P.; Neuber, E.; Eng, L. M.; White, J. S.; Rønnow, H. M.; Dewhurst, C. D.; Mochizuki, M.; Yanai, K. et al. Néel-Type Skyrmion Lattice with Confined Orientation in the Polar Magnetic Semiconductor GaV$_4$S$_8$. *Nat. Mater.* **2015**, *14*, 1116–1122.

(16) Tokunaga, Y.; Yu, X. Z.; White, J. S.; Rønnow, H. M.; Morikawa, D.; Taguchi, Y.; Tokura, Y. A New Class of Chiral Materials Hosting Magnetic Skyrmions Beyond Room Temperature. *Nat. Commun.* **2015**, *6*, 1–7.

(17) Kurumaji, T.; Nakajima, T.; Ukleev, V.; Feoktystov, A.; Arima, T. H.; Kakurai, K.; Tokura, Y. Néel-Type Skyrmion Lattice in the Tetragonal Polar Magnet VOSe$_2$O$_5$. *Phys. Rev. Lett.* **2017**, *119*, 237201.

(18) Omrani, A. A.; White, J. S.; Prša, K.; Živković, I.; Berger, H.; Magrez, A.; Liu, Y. H.; Han, J. H.; Rønnow, H. M. Exploration of the Helimagnetic and Skyrmion Lattice Phase diagram in Cu$_2$OSeO$_3$ using magnetoelectric susceptibility. *Phys. Rev. B* **2014**, *89*, 064406.

(19) Seki, S.; Ishiwata, S.; Tokura, Y. Magnetoelectric Nature of Skyrmions in a Chiral Magnetic Insulator Cu$_2$OSeO$_3$. *Phys. Rev. B* **2012**, *86*, 060403.

(20) Seki, S.; Kim, J. H.; Inosov, D. S.; Georgii, R.; Keimer, B.; Ishiwata, S.; Tokura, Y. Formation and Rotation of Skyrmion Crystal in the Chiral-Lattice Insulator Cu$_2$OSeO$_3$. *Phys. Rev. B* **2012**, *85*, 220406.

(21) White, J.S.; Prša, K.; Huang, P.; Omrani, A. A.; Živković, I.; Bartkowiak, M.; Berger, H.; Magrez, A.; Gavilano, J.L.; Nagy, G. et al. Electric-Field-Induced Skyrmion Distortion and Giant Lattice Rotation in the Magnetoelectric Insulator Cu$_2$OSeO$_3$. *Phys. Rev. Lett.* **2014**, *113*, 107203.

(22) Okamura, Y.; Yamasaki, Y.; Morikawa, D.; Honda, T.; Ukleev, V.; Nakao, H.; Murakami, Y.; Shibata, K.; Kagawa, F.; Seki, S. et al. Directional Electric-Field Induced Transformation from Skyrmion Lattice to Distinct Helices in Multiferroic Cu$_2$OSeO$_3$. *Phys. Rev. B* **2017**, *95*, 184411.

(23) Kruchkov, A. J.; White, J. S.; Bartkowiak, M.; Živković, I.; Magrez, A.; Rønnow, H. M. Direct Electric Field Control of the Skyrmion Phase in a Magnetoelectric Insulator. *Sci. Rep.* **2018**, *8*, 1–7.

(24) White, J.S.; Živković, I.; Kruchkov, A.J.; Bartkowiak, M.; Magrez, A.; Rønnow, H. M. Electric-Field-Driven Topological Phase Switching and Skyrmion-Lattice Metastability in Magnetoelectric Cu$_2$OSeO$_3$. *Phys. Rev. Appl.* **2018**, *10*, 014021.

(25) Das, B.; Balasubramanian, B.; Skomski, R.; Mukherjee, P.; Valloppilly, S. R.; Hadjipanayis, G. C.; Sellmyer, D.J. Effect of Size Confinement on Skyrmionic Properties of MnSi Nanomagnets. *Nanoscale* **2018**, *10*, 9504–9508.

(26) Devi, M.M.; Koshibae, W.; Sharma, G.; Tomar, R.; Gaikwad, V.M.; Varma, R. M.; Nair, M. N.; Jha, M.; Sarma, D. D., Chatterjee, R. et al. The Limit to Realize an Isolated Magnetic Single Skyrmionic State. *J. Mater. Chem. C* **2019**, *7*, 1337–1344.

(27) Niitsu, K.; Liu, Y.; Booth, A.; Yu, X.; Mathur, N.; Stolt, M.; Shindo, D.; Jin, S.; Zang, J.; Nagaosa, N. et al. Geometrically Stabilized Skyrmionic Vortex in FeGe Tetrahedral Nanoparticles. *Nat. Mater.* **2022**, *21*, 305-310.

(28) Du, H.; Che, R.; Kong, L.; Zhao, X.; Jin, C.; Wang, C.; Yang, J.; Ning, W.; Li, R.; Jin, C. et al. Edge-Mediated Skyrmion Chain and its Collective Dynamics in a Confined Geometry. *Nat. Commun.* **2015**, *6*, 1–7.

(29) Du, H.; Liang, D.; Jin, C.; Kong, L.; Stolt, M. J.; Ning, W.; Yang, J.; Xing, Y.; Wang, J.; Che, R. et al. Electrical Probing of Field-Driven Cascading Quantized Transitions of Skyrmion Cluster States in MnSi Nanowires. *Nat. Commun.* **2015**, *6*, 1–7.

(30) Yu, X. Z.; DeGrave, J. P.; Hara, Y.; Hara, T.; Jin, S.; Tokura, Y. Observation of the Magnetic Skyrmion Lattice in a MnSi Nanowire by Lorentz TEM. *Nano Lett.* **2013**, *13*, 3755.

(31) Zheng, F.; Li, H.; Wang, S.; Song, D.; Jin, C.; Wei, W.; Kovács, A.; Zang, J.; Tian, M.; Zhang, Y. et al. Direct Imaging of a Zero-Field Target Skyrmion and Its Polarity Switch in a Chiral Magnetic Nanodisk. *Phys. Rev. Lett.* **2017**, *119*, 197205.

(32) Tang, J.; Wu, Y.; Kong, L.; Wang, W.; Chen, Y.; Wang, Y.; Soh, Y.; Xiong, Y.; Tian, M.; Du, H. Two-Dimensional Characterization of Three-Dimensional Magnetic Bubbles in $Fe_3Sn_2$ Nanostructures. *Natl Sci. Rev.* **2021**, *8*, nwaa200.

(33) Jin, C.; Li, Z.A.; Kovács, A.; Caron, J.; Zheng, F.; Rybakov F.N.; Kiselev N.S.; Du, H.; Blügel, S.; Tian, M. et al. Control of Morphology and Formation of Highly Geometrically Confined Magnetic Skyrmions. *Nat. Commun.* **2017**, *8*, pp.1-9.

(34) Ni, Y.; Li, H.; Jin, L.; Hong, J. Synthesis of 1D $Cu(OH)_2$ Nanowires and Transition to 3D CuO Microstructures under Ultrasonic Irradiation, and Their Electrochemical Property. *Cryst. Growth Des.* **2009**, *9*, 3868–3873.

(35) Dyadkin, V.; Prša, K.; Grigoriev, S. V.; White, J.S.; Huang, P.; Rønnow, H. M.; Magrez, A.; Dewhurst, C. D.; Chernyshov, D. Chirality of Structure and Magnetism in the Magnetoelectric Compound $Cu_2OSeO_3$. *Phys. Rev. B* **2014**, *89*, 140409.

(36) Baral, P. R.; Ukleev, V.; Cubitt, R.; Yu, L.; Rønnow, H. M.; White, J. S.; Magrez, A. Institut Laue-Langevin (ILL) DOI:10.5291/ILL-DATA.5-31-2696; **2020**.

(37) Dewhurst, C. D. GRASP User Manual, *Technical Report No. ILL03DE01T, Institut Laue- Langevin, Grenoble* **2003**, available at https://www.ill.eu/users/support-labs-infrastructure/software-scientific-tools/grasp/

(38) Vansteenkiste, A.; Leliaert, J.; Dvornik, M.; Helsen, M.; Garcia-Sanchez, F.; Van Waeyenberge, B. The Design and Verification of MuMax3. *AIP advances* **2014**, *4*, 107133.

(39) Zhang, S.; Van der Laan, G.; Müller, J.; Heinen, L.; Garst, M.; Bauer, A.; Berger, H.; Pfleiderer, C.; Hesjedal, T. Reciprocal Space Tomography of 3D Skyrmion Lattice Order in a Chiral Magnet. *Proc. Natl. Acad. Sci. U.S.A.* **2018**, *115*, 6386–6391.

(40) Janson, O.; Rousochatzakis, I.; Tsirlin, A. A.; Belesi, M.; Leonov, A. A.; Rößler, U. K.; Van Den Brink, J.; Rosner, H. The Quantum Nature of Skyrmions and Half-Skyrmions in $Cu_2OSeO_3$. *Nat. Commun.* **2014**, *5*, 1–11.

(41) Müller, G. P.; Hoffmann, M.; Dißelkamp, C.; Schürhoff, D.; Mavros, S.; Sallermann, M.; Kiselev, N. S.; Jónsson, H.; Blügel, S. Spirit : Multifunctional Framework for Atomistic Spin Simulations. *Phys. Rev. B* **2019**, *99*, 224414.

(42) Ahrens, J., Geveci, B., Law, C. ParaView: An End-User Tool for Large Data Visualization,Visualization Handbook; Elsevier: 2005, ISBN- 13: 978-0123875822; see also: http://www.paraview.org (accessed March 12, 2021).

(43) Adams, T.; Chacon, A.; Wagner, M.; Bauer, A.; Brandl, G.; Pedersen, B.; Berger, H.; Lemmens, P.; Pfleiderer, C. Long-Wavelength Helimagnetic Order and Skyrmion Lattice Phase in $Cu_2OSeO_3$. *Phys. Rev. Lett.* **2012**, *108*, 237204.

(44) White, J. S.; Levatić, I.; Omrani, A. A., Egetenmeyer, N., Prša, K., Živković, I., Gavilano, J. L., Kohlbrecher, J., Bartkowiak, et al. Electric Field Control of the Skyrmion Lattice in $Cu_2OSeO_3$. *J. Condens. Matter Phys.* **2012**, *24*, 432201.

(45) Bauer, A.; Pfleiderer, C. Magnetic Phase Diagram of MnSi Inferred From Magnetization and ac-Susceptibility. *Phys. Rev. B* **2012**, *8*, 214418.

(46) Wilhelm, H.; Baenitz, M.; Schmidt, M.; Rößler, U. K.; Leonov, A. A.; Bogdanov, A.N. Precursor Phenomena at the Magnetic Ordering of the Cubic Helimagnet FeGe. *Phys. Rev. Lett.* **2011**, *107*, 127203.

(47) Bauer, A.; Garst, M.; Pfleiderer, C. History Dependence of the Magnetic Properties of Single-Crystal $Fe_{1-x}Co_xSi$. *Phys. Rev. B* **2016**, *93*, 235144.

(48) Qian, F.; Wilhelm, H.; Aqeel, A.; Palstra, T. T. M.; Lefering, A. J. E.; Brück, E. H.; Pappas, C. Phase Diagram and Magnetic Relaxation Phenomena in $Cu_2OSeO_3$. *Phys. Rev. B* **2016**, *94*, 064418.

(49) Leonov, A.O.; Togawa, Y.; Monchesky, T.L.; Bogdanov, A.N.; Kishine, J.; Kousaka, Y.; Miyagawa, M.; Koyama, T.; Akimitsu, J.; Koyama, T. et al. Chiral Surface Twists and Skyrmion Stability in Nanolayers of Cubic Helimagnets. *Phys. Rev. Lett.* **2016**, *117*, 087202.

(50) Živković, I.; White, J. S.; Rønnow, H. M.; Prša, K.; Berger, H. Critical Scaling in the Cubic Helimagnet $Cu_2OSeO_3$. *Phys. Rev. B* **2014**, *89*, 060401.

(51) Kindervater, J.; Adams, T.; Bauer, A.; Haslbeck, F. X.; Chacon, A.; Mühlbauer, S.; Jonietz, F.; Neubauer, A.; Gasser, U.; Nagy, G. et al. Evolution of Magnetocrystalline Anisotropies in $Mn_{1-x}Fe_xSi$ and $Mn_{1-x}Co_xSi$ as Inferred from Small-Angle Neutron Scattering and Bulk Properties. *Phys. Rev. B* **2020**, *101*, 104406.

(52) Ukleev, V.; Utesov, O.; Yu, L.; Luo, C.; Chen, K.; Radu, F.; Yamasaki, Y.; Kanazawa, N.; Tokura, Y. et al. Signature of Anisotropic Exchange Interaction Revealed by Vector-Field Control of the Helical Order in a FeGe Thin Plate. *Phys. Rev. Res.* **2021**, *3*, 013094.

(53) Karube, K.; White, J. S.; Reynolds, N.; Gavilano, J. L.; Oike, H.; Kikkawa, A.; Kagawa, F.; Tokunaga, Y.; Rønnow, H. M.; Tokura, Y. et al. Robust Metastable Skyrmions and their Triangular–Square Lattice Structural Transition in a High-Temperature Chiral Magnet. *Nat. Mater.* **2016**, *15*, 1237–1242.

(54) Chacon, A.; Heinen, L.; Halder, M.; Bauer, A.; Simeth, W.; Mühlbauer, S.; Berger, H.; Garst, M.; Rosch, A.; Pfleiderer, C. Observation of Two Independent Skyrmion Phases in a Chiral Magnetic Material. *Nat. Phys.* **2018**, *14*, 936–941.

(55) Pathak, S. A.; Hertel, R. Three-Dimensional Chiral Magnetization Structures in FeGe Nanospheres. *Phys. Rev. B* **2021**, *103*, 104414.

(56) Rybakov, F. N.; Borisov, A.B.; Blügel, S.; Kiselev, N.S. New Type of Stable Particle like States in Chiral Magnets. *Phys. Rev. Lett.* **2015**, *115*, 117201.

(57) Zheng, F.; Rybakov, F. N.; Borisov, A.B.; Song, D.; Wang, S.; Li, Z.-A.; Du, H.; Kiselev, N.S.; Caron, J.; Kovacs, A. et al. Experimental Observation of Chiral Magnetic Bobbers in B20-type FeGe. *Nature Nanotech.* **2018**, *13*, 451-455.

(58) Ahmed, A.; Rowland, J.; Esser, B.D.; Dunsiger, S.R.; McComb, D.W.; Randeria, M.; Kawakami, R.K. Chiral Bobbers and Skyrmions in Epitaxial FeGe/Si(111) Films. *Phys. Rev. Mat.* **2018**, *2*, 041401.

(59) Ran,K.; Liu, Y.; Guang, Y.; Burn, D.M.; van der Laan, G.; Hesjedal, T.; Du, H.; Yu,G.; Zhang, S. Creation of a Chiral Bobber Lattice in Helimagnet-Multilayer Heterostructures. *Phys. Rev. Lett.* **2021**, *126*, 017204.

(60) Chui, C. P.; Ma, F.; Zhou, Y. Geometrical and Physical Conditions for Skyrmion Stability in a Nanowire. *AIP Advances* **2015**, *5*, 047141.

(61) Pepper, R. A.; Beg, M.; Cortés-Ortuño, D.; Kluyver, T.; Bisotti, M.A.; Carey, R.; Vousden, M.; Albert, M.; Wang, W.; Hovorka, O. et al. Skyrmion States in Thin Confined Polygonal Nanostructures. *J. Appl. Phys.* **2018**, *123*, 093903.

(62) Donnelly, C.; Scagnoli, V. Imaging Three-Dimensional Magnetic Systems with X-Rays. *J. Phys.: Condens. Matter* **2020**, *32*, 213001.

(63) Heacock, B.; Sarenac, D.; Cory, D. G.; Huber, M. G.; MacLean, J. P. W.; Miao, H.; Wen, H.; Pushin, D. A. Neutron Sub-Micrometre Tomography from Scattering Data, *IUCrJ* **2020**, *5*, 893.

(64) Wilson, M. N.; Karhu, E. A.; Lake, D. P.; Quigley, A. S.; Meynell, S.; Bogdanov, A. N.; Fritzsche, H.; Rößler, U. K.; Monchesky, T. L. Discrete Helicoidal States in Chiral Magnetic Thin Films. *Phys. Rev. B* **2013**, *88*, 214420.

(65) Karhu, E.A.; Kahwaji, S.; Robertson, M.D.; Fritzsche, H.; Kirby, B.J.; Majkrzak, C.F.; Monchesky, T.L. Helical Magnetic Order in MnSi Thin Films. *Phys. Rev. B* **2011**, *84*, 060404.

(66) Kanazawa, N.; White, J. S.; Rønnow, H. M.; Dewhurst, C. D.; Fujishiro, Y.; Tsukazaki, A.; Kozuka, Y.; Kawasaki, M.; Ichikawa, M.; Kagawa, F. et al. Direct Observation of Anisotropic Magnetic Field Response of the Spin Helix in FeGe Thin Films. *Phys. Rev. B* **2016**, *94*, 184432.

(67) Porter, N. A.; Spencer, C. S.; Temple, R. C.; Kinane, C. J.; Charlton, T. R.; Langridge, S.; Marrows, C. H. Manipulation of the Spin Helix in FeGe Thin Films and FeGe/Fe Multilayers. *Phys. Rev. B* **2015**, *92*, 144402.

(68) Open access to the dataset and code used for this article. DOI: 10.5281/zenodo.6651369; **2022**.

**TOC GRAPHIC**

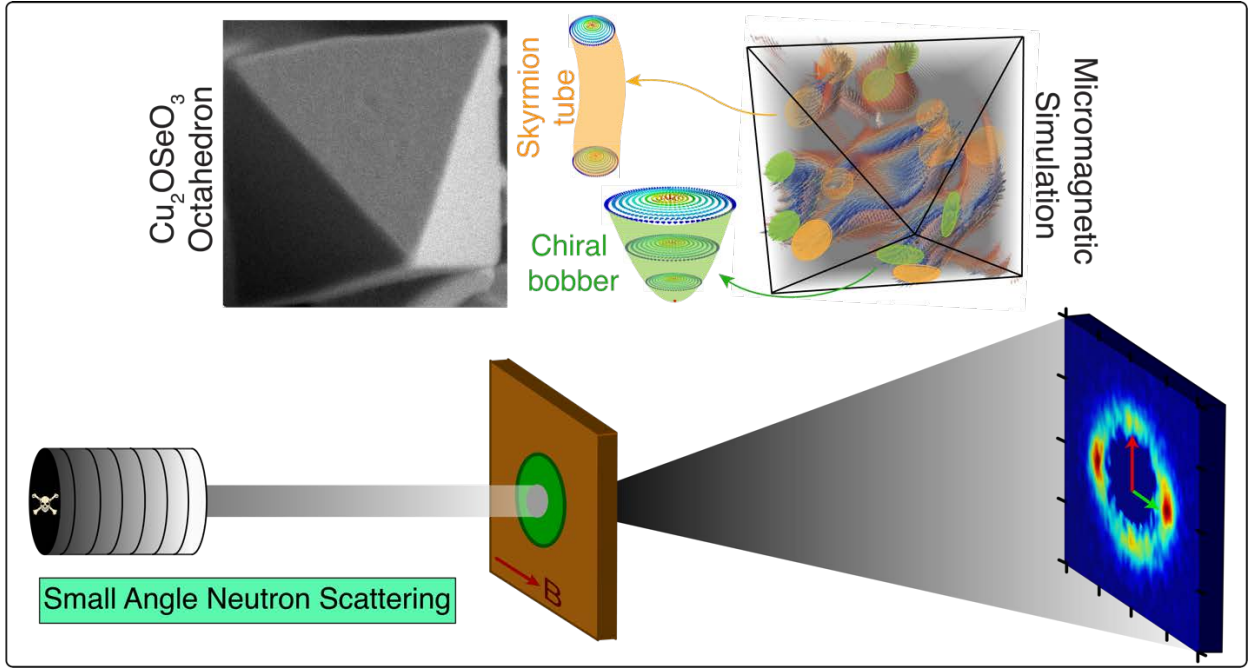

# Supporting information for

# "Tuning Topological Spin Textures in Size-Tailored Chiral Magnet Insulator Particles"

*Priya R. Baral[a,b], Victor Ukleev[c], Thomas LaGrange[d], Robert Cubitt[e], Ivica Živković[f], Henrik M. Rønnow[f], Jonathan S. White[c], Arnaud Magrez[a,*]*

a) Crystal Growth Facility, Institute of Physics, École Polytechnique Fédérale de Lausanne (EPFL), CH-1015 Lausanne, Switzerland

b) Chair of Computational Condensed Matter Physics, Institute of Physics, École Polytechnique Fédérale de Lausanne (EPFL), CH-1015 Lausanne, Switzerland

c) Laboratory for Neutron Scattering and Imaging (LNS), Paul Scherrer Institut (PSI), CH-5232 Villigen, Switzerland

d) Laboratory for Ultrafast Microscopy and Electron Scattering (LUMES), Institute of Physics, École Polytechnique Fédérale de Lausanne (EPFL), CH-1015 Lausanne, Switzerland

e) Institut Laue Langevin, Large Scale Structures, 71 Avenue des Martyrs CS 20156, 38042 Grenoble, France

f) Laboratory for Quantum Magnetism, Institute of Physics, École Polytechnique Fédérale de Lausanne (EPFL), CH-1015 Lausanne, Switzerland

* Email : arnaud.magrez@epfl.ch

**Chemistry of Cu$_2$OSeO$_3$ particles.** In order to validate the mechanism of conversion of CuSeO$_3$·2H$_2$O into Cu$_2$OSeO$_3$, high temporal resolution (0.5 second for each pattern) XRD measurements were performed at SNBL, ESRF (Fig. S1A).

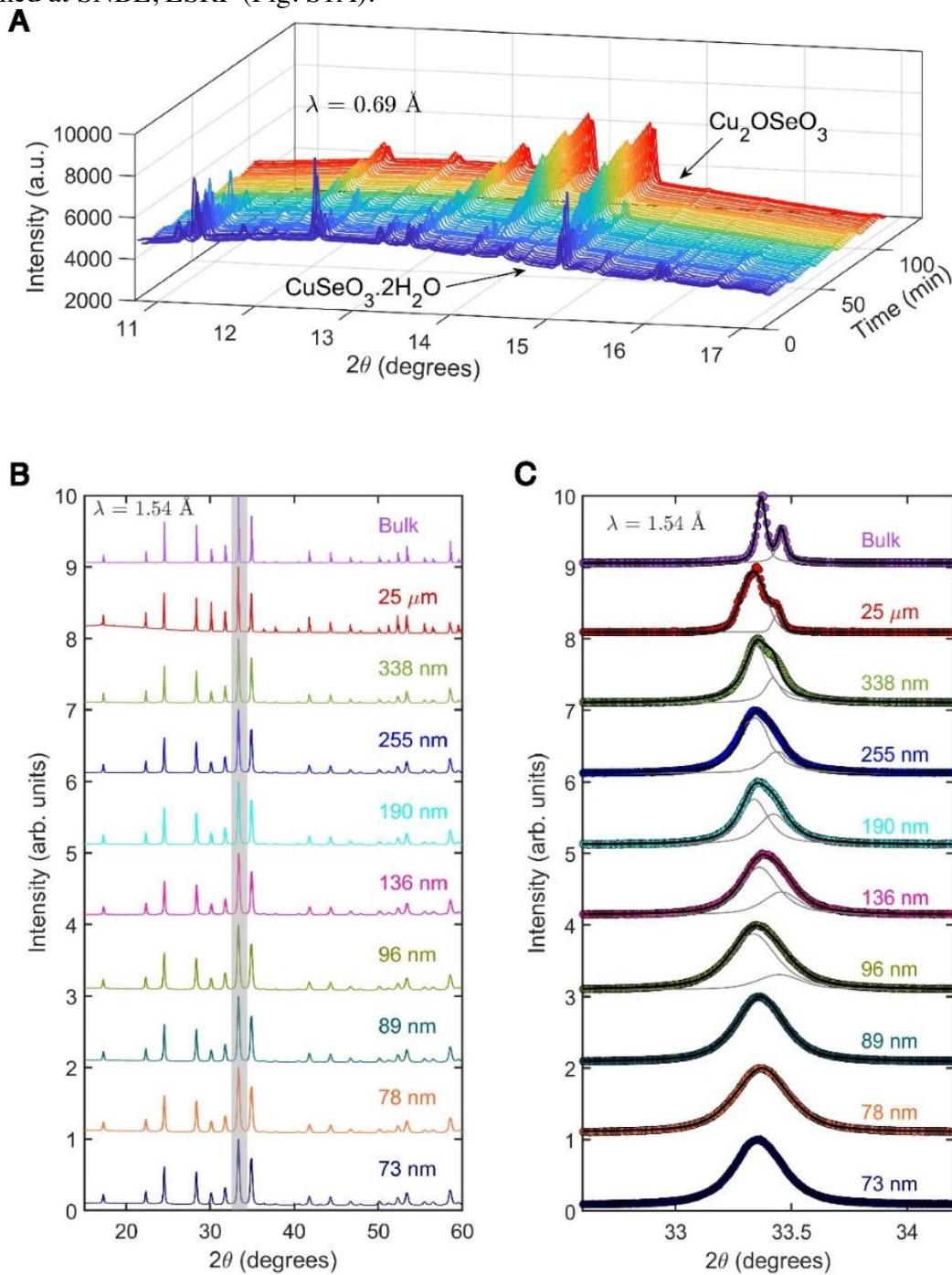

**Fig. S1.** (A) High resolution synchrotron XRD patterns of *in situ* transformation between the precursor CuSeO$_3$·2H$_2$O and final product Cu$_2$OSeO$_3$, performed at SNBL, ESRF. This solid-solid transformation excludes the possibility of having any impurity phases throughout the whole process including the final Cu$_2$OSeO$_3$ phase. (B) High-resolution room temperature XRD patterns of all the particles, compared with that of the bulk sample. All specimens appear to be phase pure. (C) From the complete $2\theta$ range, one peak at $2\theta \sim 33.4$ degrees has been singled out for all the particles and fitted with two Pseudo-Voigt functions for Cu-K$_{\alpha1}$ and Cu-K$_{\alpha2}$. In the bulk sample two independent functions are needed in order to have a reasonable fit. Whereas for the smallest particles, single Pseudo-Voigt function is sufficient in order to have an adequate fit, displaying the direct consequences of size broadening on the diffraction peaks.

In the presence of a basic medium as $NH_4OH$, $CuSeO_3.2H_2O$ transforms completely to $Cu_2OSeO_3$, with no trace of any other impurity phase(s). It provides us with great insight into the microscopic mechanism of formation of the final product. But this is beyond the scope of this work and will be reported somewhere else. The first insight into the size control of the particles is evident from the room-temperature XRD patterns (Fig. S1B & C). The evidence of peak broadening is obvious from here. In theory, this peak broadening can be a result of various factors, such as dislocations, twinning, coherency strain, chemical heterogeneity and, finally, small crystallite sizes. If these factors, other than the last one, were responsible for the observations, it would also be evident from other local characterization tools (Fig. 2E & F of the main text). Moreover, presence of atomistic crystalline defects may prevent the formation of a skyrmion, a highly symmetric object. And finally, since the instrumental contribution was fixed for all the patterns, the broadening of diffraction peaks is thus attributed to average coherent size of the particles.

**Particle size determination.** In our work, the particle size was determined by using a combination of dynamic light scattering (DLS), SEM imaging and Williamson-Hall analysis. As shown in the SEM image below, the as-grown sample is typically synthesized as agglomerates of well faceted and highly crystalline particles. Thus, the whole procedure was divided into two steps. In the first step, the secondary particle size was determined by DLS technique. The powder sample was suspended in a water + surfactant solution (SDS - Sodium Dodecyl Sulfate), followed by thorough sonication. The measurement was performed in a Zetasizer and for consistency check, 10-15 repetitions were performed for each sample size. As can be seen from the figure below, the DLS technique correctly measures the secondary particle sizes, for all samples. The example shown below corresponds to the one with mean size of 1324 ± 5 nm. The measurements confirm the very narrow distribution of the secondary particle size as opposed to the distribution reported so far in Holt *et al.*, Mater. Res. Express 8 116101 (2021).

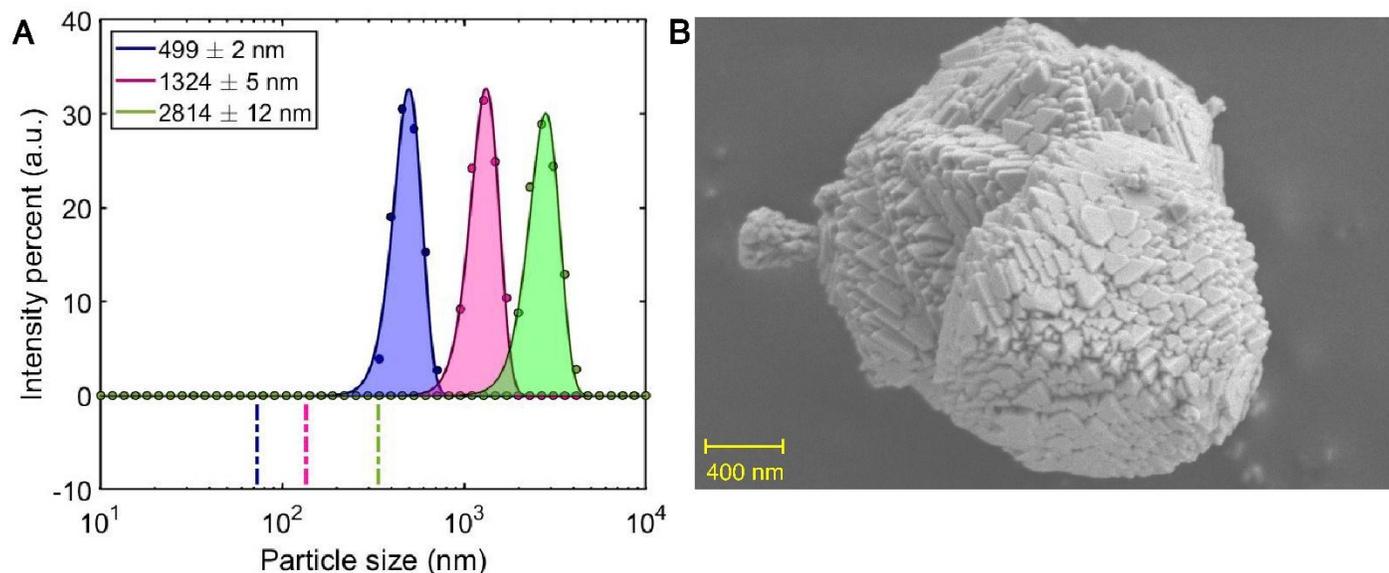

**Fig. S2.** (A) Particle size determination from DLS measurements. The maximum of the shaded region corresponds to the average secondary size as determined from the experiments. The dashed lines are the average primary particle sizes, as obtained from W-H analysis (shown in Fig. 2E of the main text). (B) An example SEM image representing both the primary as well as secondary particle sizes for the batch labeled in magenta in Panel-A.

But from the perspective of skyrmion physics, size of the primary particles is much more relevant compared to the secondary. Although the distribution appears to be narrow in panel B, an accurate particle size measurement cannot be obtained from the SEM images because particles are tilted, overlapping and/or partially embedded in the agglomerates. So, 3D particle size distribution cannot not be obtained from SEM. We preferred to determine the primary particle size by performing Williamson-Hall (W-H) analysis. A large 2θ range is used such that the 3D particle size can be accurately obtained. As shown in the Fig. 2, the primary particle size corresponding to the sample in the Panel B image is determined to be 136 ± 2 nm. Together with all the measurement protocols described above, we firmly believe in a tight distribution of the particle sizes, which correlates directly with the systematic behavior observed in the experimental magnetic properties.

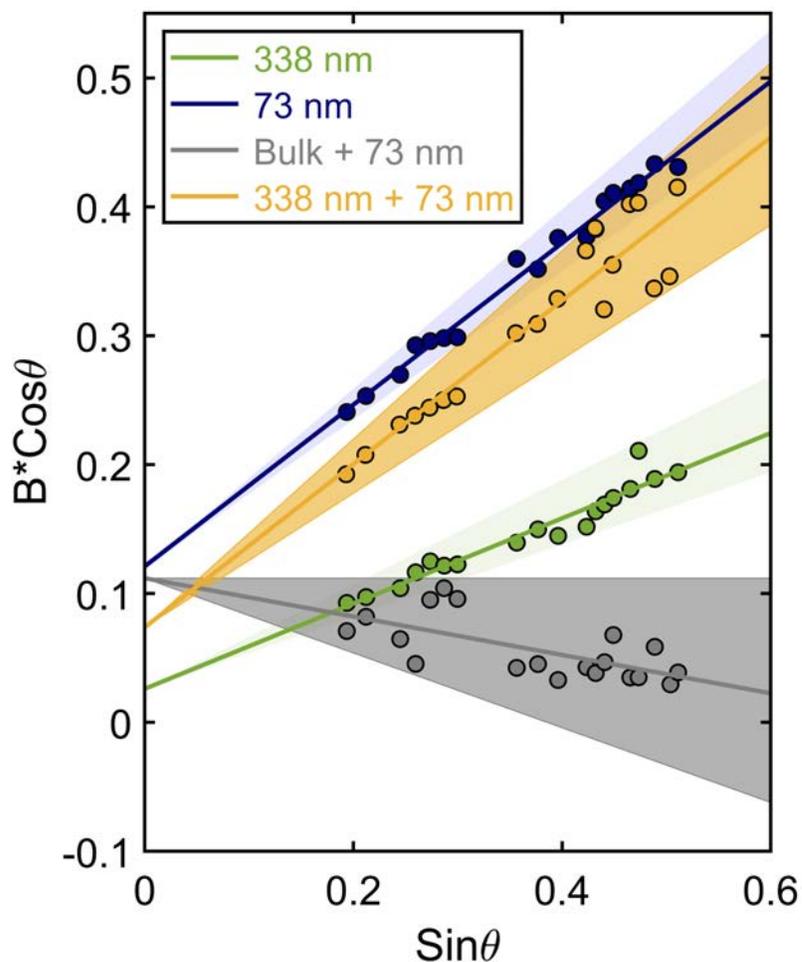

**Fig. S3.** Williamson-Hall measurements performed on a mixture of two particle sizes. Data for 338 nm and 73 nm have been shown for reference. The procedure to obtain the average particle height is identical to the ones described in the methods section.

In order to confirm the narrow size distribution, we performed some additional W-H measurements on a mixture made of 73 nm particles with 338 nm particles such that the size distribution becomes broad. Figure S3 shows that while the W-H plot for pure 73 nm and 338 nm samples are perfectly linear, a strong scattering is obtained at high 2θ angles for the mixture of 73 nm and 338 nm. Same broadening in the W-H plot is seen when the 73 nm particles are mixed with a powder obtained by crushing a $Cu_2OSeO_3$ single crystal. These additional measurements confirm the primary particle size distribution is narrow.

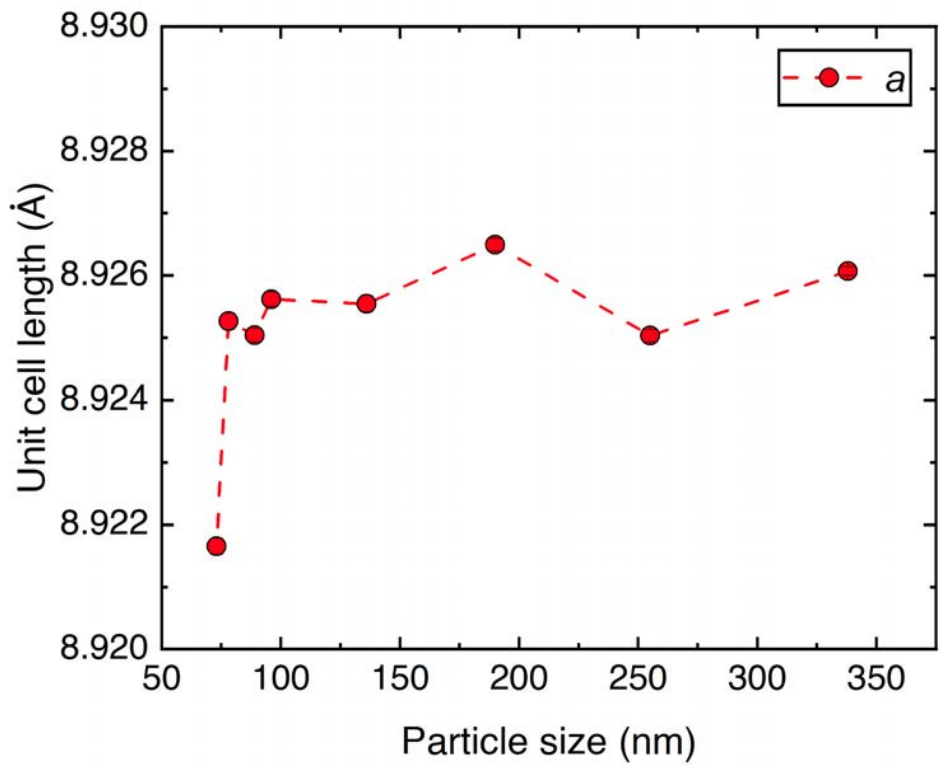

**Fig. S4.** Refinement of the powder patterns used for WH analysis. The obtained length of the unit cell lies very close to the reported value of 8.925 Å.

**Transmission Electron Microscopy Studies.** The crystallinity and structure of $Cu_2OSeO_3$ nanoparticles were investigated in selected area electron and high-resolution transmission electron microscopy (HRTEM). The as-fabricated $Cu_2OSeO_3$ nanoparticles were put into a solution of ethanol. The solution was then sonicated to break the aggregated nanoparticles and then using a micropipette, the solution of nanoparticles was drop cast onto TEM lacy carbon grids. The nanoparticles have a morphology of highly faceted cubic particles (as observed by SEM). Figure S5 (A) shows an example of such $Cu_2OSeO_3$ nanoparticle. Diffraction analysis was conducted on several particles, and investigation confirmed the high crystallinity and that particles were single crystals having $Cu_2OSeO_3$ structure. Fig. S5 shows the diffraction of a faceted cubic-shaped particle. The sample was tilted to the closest zone axis [311]. The diffraction pattern was simulated in the JEMS software[1]. The calculation includes dynamical scattering effects for a 400 nm particle and the microscope coherence and electron beam illumination conditions. The JEMS simulation of selected areas electron diffraction pattern in Figure S5B shows excellent agreement for the $Cu_2OSeO_3$ crystal oriented on [311] zone axis.

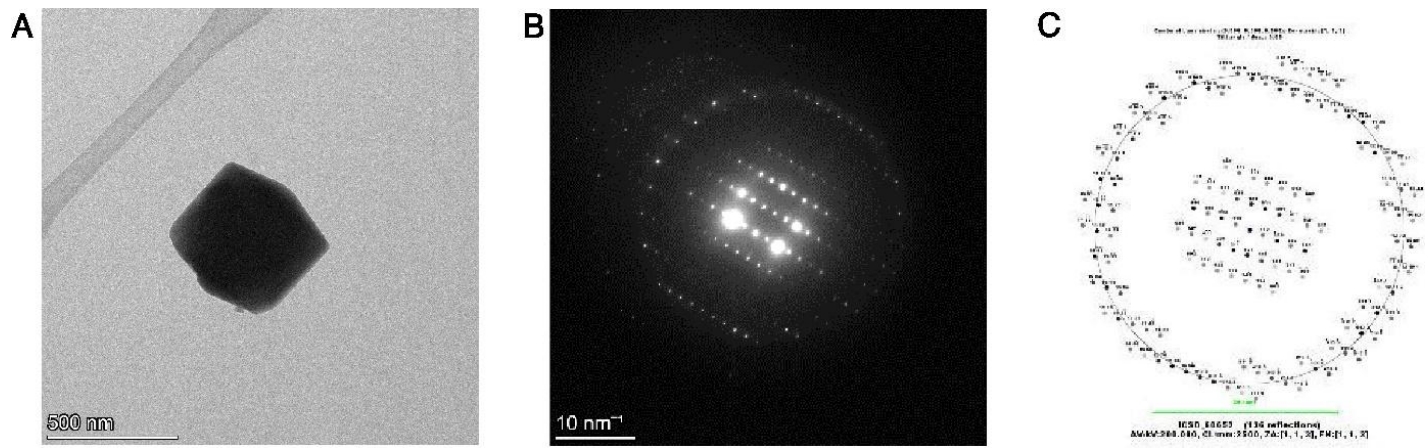

**Fig. S5.** (A) TEM image of faceted Cu$_2$OSeO$_3$ nanoparticle. (B) Experimental Selected Area Diffraction Pattern (SAED). (C) JEMS simulation of SAED of [311] zone axis that includes zero and first order Laue zones.

    High-resolution transmission electron microscopy (HRTEM) investigations were conducted using Thermo Fisher Scientific Talos TEM at 200kV accelerating voltage. Figure S6 shows HRTEM images taken from the edge region of the cubic shape Cu$_2$OSeO$_3$ nanoparticle shown in Fig. S5 (A) tilted on [311] zone axis. We used the JEMS software package and performed a multi-slice simulation of the HRTEM image for 50 nm thick crystal on [311] zone axis and the Talos microscope conditions were used to acquire images in Fig. S6 (A) and (B). The HRTEM simulation shows good agreement with the experimental HRTEM for defocus of 67 nm close to the Scherzer defocus (optimal) based on the aberration and coherence specifications of the Thermo Fisher Scientific Talos TEM. The SAED and HRTEM investigations confirm that Cu$_2$OSeO$_3$ nanoparticles are single crystals and do not have apparent defects.

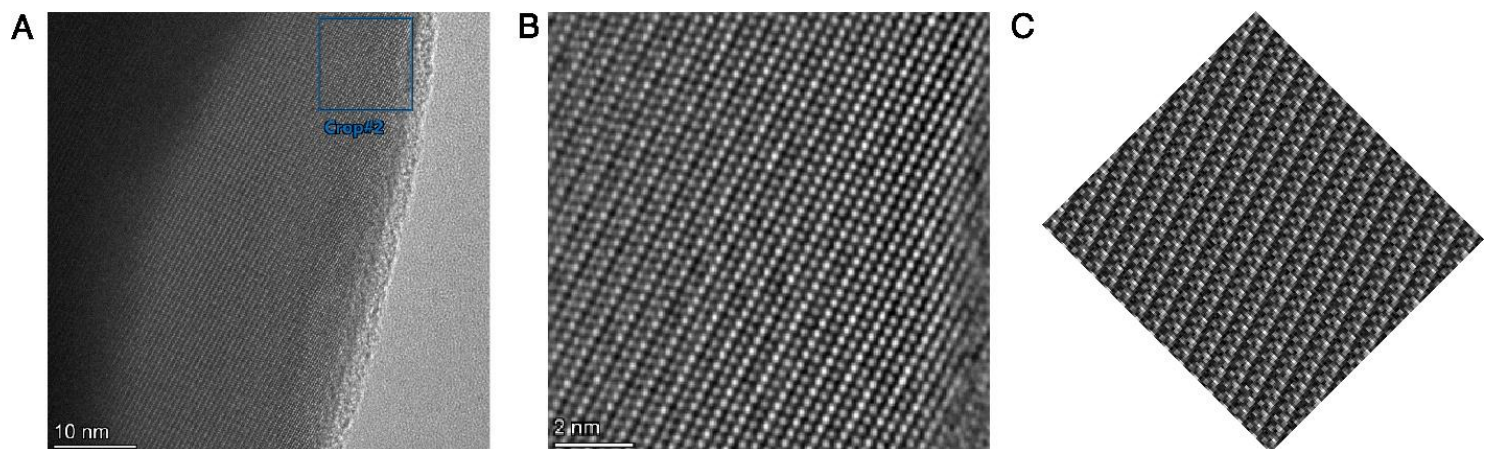

**Fig. S6.** (A) HRTEM image of the edge of Cu$_2$OSeO$_3$ nanoparticle in Fig. S5 (A) titled to be aligned on [311] zone axis. (B) higher magnification HRTEM image of area indicated by the blue in Fig S5 (A) showing atomic resolved lattice planes. (C) JEMS multi-slice calculation simulation of the Cu$_2$OSeO$_3$ nanoparticle with [311] orientation and 50 nm crystal thickness.

**Magnetometry.** AC susceptibility field scans were performed for sample of all particle sizes at constant temperatures (Fig. S7A-G), in a similar manner to those shown in Fig. 3C, F, and I. Since the contour plots are created out of the imaginary component of the total susceptibility, a rather growing dissipative region can be observed close to $T_c$, for reduced particle size. Motivated by this, AC susceptibility temperature scans were performed at constant magnetic fields as shown in Fig. S8. Strong correlations among the chiral fluctuations close to $T_c$, drive the HM-transition to lower temperatures in bulk-$Cu_2OSeO_3$.

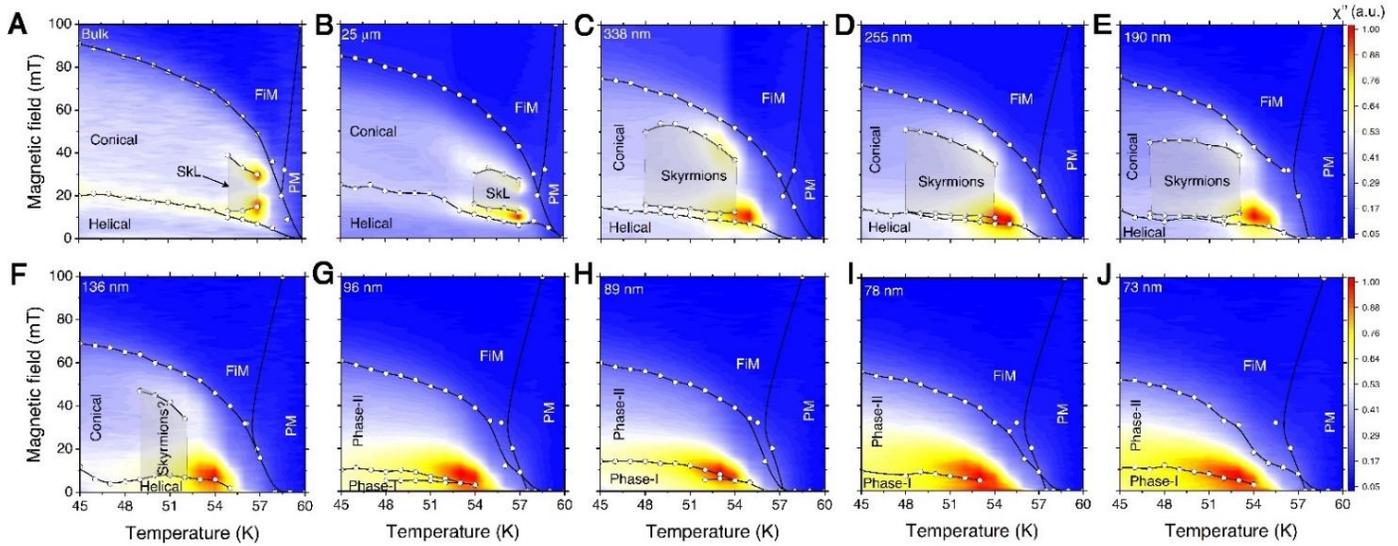

**Fig. S7.** (A)-(J) display the magnetic phase diagrams of all particle sizes considered in the text. Panel A, C and J are already shown in Fig.3 of the main text. The phase boundaries have been extracted from the AC susceptibility measurements, following similar protocols as shown in Fig. 3B, E and H of the main text.

The divergence in the susceptibility close to $T_c$ for smaller particles signifies a direct consequence of finite-size effects. Similar results were obtained with 100 mT of applied magnetic field (Fig. S8E), suggesting the applicability of these finite-size effects not only in zero-field transitions, but also finite-field crossovers. In bulk-$Cu_2OSeO_3$, temperature-cooling scan in presence of 20 mT magnetic field drives the system through PM → SkL → Con. phase, represented by the cusp in the susceptibility data (Fig. S8D). In the absence of a well-defined lattice of skyrmions, this cusp vanishes in the particles. In the biggest particles, a small change in the susceptibility can be seen. These changes are thus attributed to the presence of few skyrmion tubes and/or chiral bobbers. Whereas, in the smallest particles, absence of this feature not only disfavors the possibility of having skyrmion tubes but also completely agrees with the SANS observations.

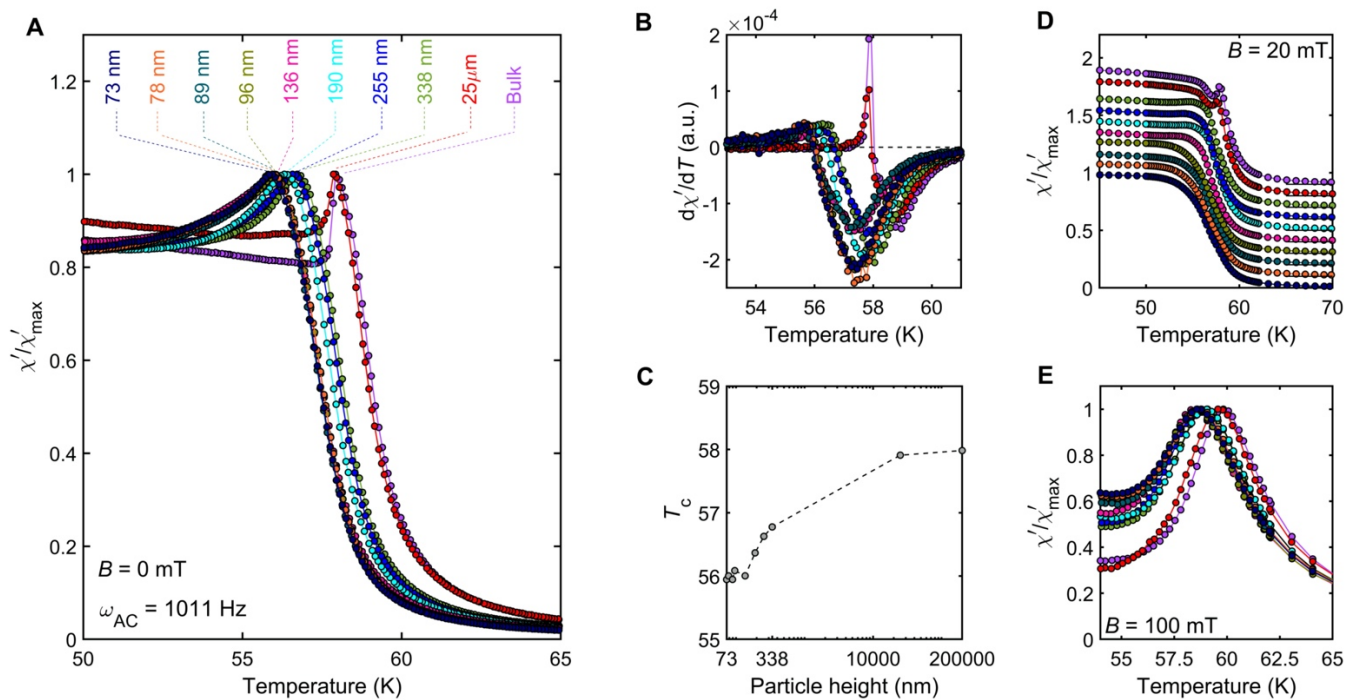

**Fig. S8.** (A) Zero-field AC susceptibility temperature scans for all the particles. In order to emphasize the fact that $T_c$ decreases with decreasing particle size, all the scans have been normalized with respect to the maximum. (B) shows the differentiation with respect to temperature for all the scans shown before, thus extracting the $T_c$. (C) The estimated $T_c$ from the previous procedure compared with all particle sizes considered. AC susceptibility temperature scan in presence of (D) 20 mT and (E) 100 mT of DC magnetic field, also for all the particles.

**Small angle neutron scattering.** SANS measurements for the biggest particles (338 nm) shed more lights on the magnetic phases established in Fig. 5A-F of the main text. In order to suppress the issue of multiple scattering, a pellet of thickness less than 0.5 mm was used in a custom-made Al-holder. Firstly at 53 K in presence of zero magnetic field, as shown in Fig. S9A, the detector image shows an isotropic scattering ring around $Q = 0$. This scattering pattern may be the result of helical spin textures within each of the randomly oriented octahedron of $Cu_2OSeO_3$. The $|q|$ of this scattering pattern, analyzed with a Pseudo-Voight function, closely matches with the one obtained previously from bulk-$Cu_2OSeO_3$. In contrast, for the smallest nanoparticles (73 nm), $|q|$ was found to be around 0.087(1) nm$^{-1}$. In real space, this closely follows the entire height of the octahedron particles, thus outlining the main effects of confinements due to reduced crystallite size. The evolution of this $|q|$ with external magnetic field suggests that the pitch length slowly relaxes to the bulk limit before vanishing in the field-polarized state.

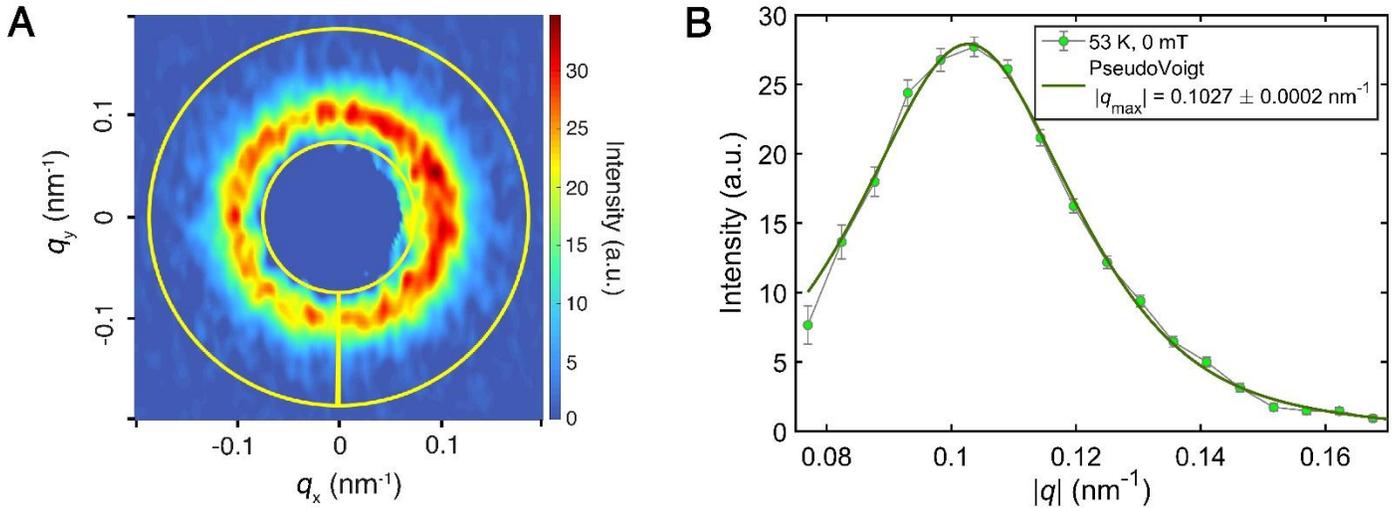

**Fig. S9.** (A) SANS detector image for 338 nm particles at 53 K at zero magnetic field. In order to estimate the total radial intensity evolution as a function of $|q|$, an isotropic sector was chosen as shown in yellow. To ensure all magnetic scattering is considered for this analysis, the width of the sector was chosen such that it spans the maximum of the detector image. (B) magnetic scattering as a function of $q$ extracted from the sector box defined in (A). The green solid line represents a Pseudo-Voight fit to the same.

We note that a single period helical spiral can also produce a "two-spot" SANS pattern, as shown in Fig. S11. Here, we demonstrate the single 73 nm helical pitch embedded into the octahedron with 73 nm height, and the corresponding two-dimensional fast Fourier transform (FFT) of the $m_z$-component of the magnetization. This FFT pattern can serve as a rough approximation of the SANS intensity map in the ($yz$) plane, where the neutron beam $k_i$ is parallel to $x$ and helical propagation axis is parallel to $y$ [D. A. Gilbert, *et al*. Physical Review Materials 3.1 (2019): 014408]. Two bright spots, corresponding to the magnetic structure factor are clearly observed in the FFT (Fig. S11B). The possible emergence of aperiodic arrangements of magnetic skyrmions, chiral bobbers, merons or other isolated textures embedded into ferromagnetic background should be manifested in form-factor small-angle scattering across the large $Q$-range [L. G. Vivas, *et al*. Physical Review Letters 125.11 (2020): 117201]. With the present dataset at hand we are not able to discuss this weak effect due to the limited dynamic range, and predominance of the peaked scattering intensity.

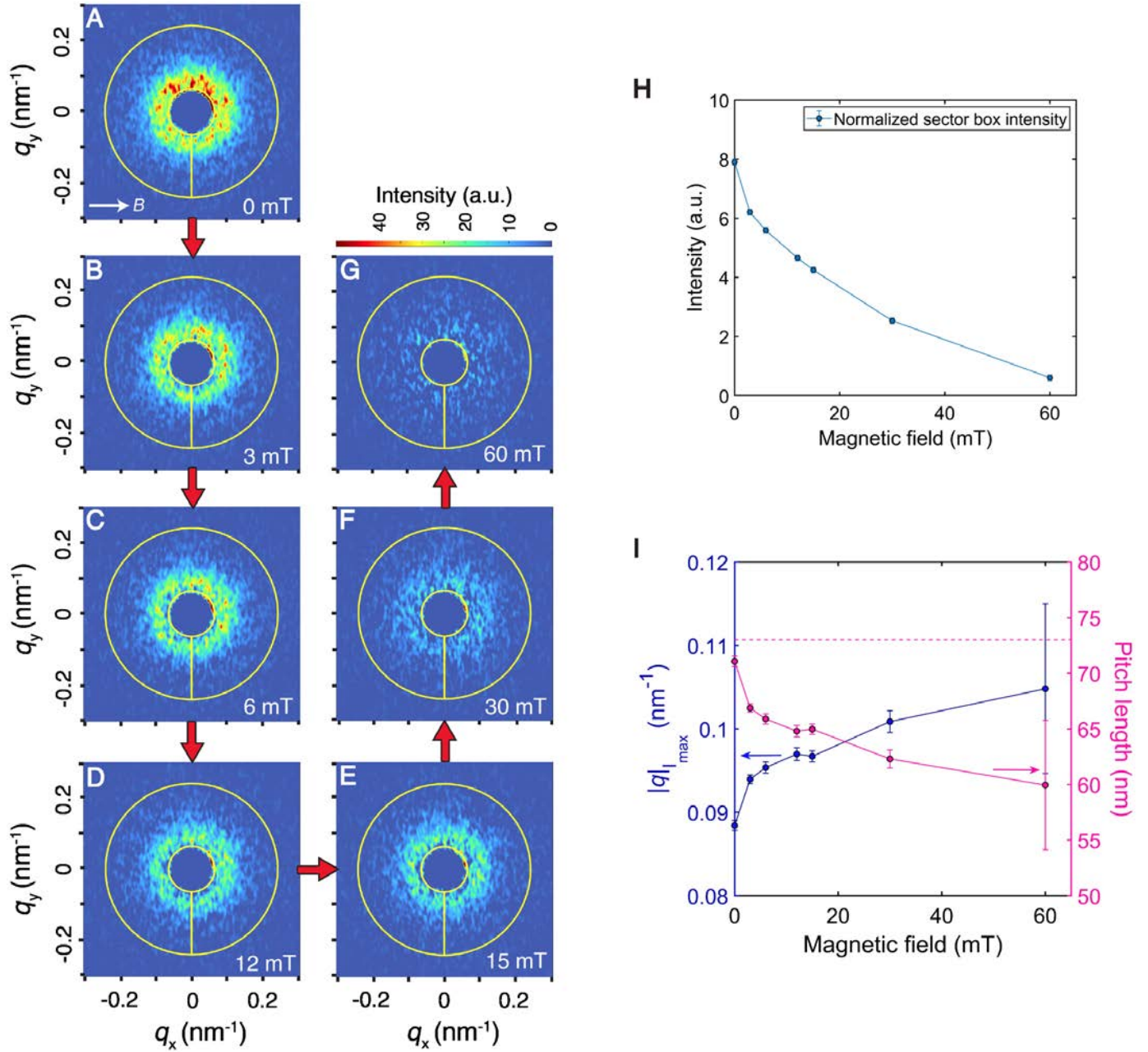

**Fig. S10.** (A)-(G) represent SANS detector images for the smallest nanoparticles (73 nm) measured in the same instrumental conditions as described previously in Fig. 4 of the main text. In order to estimate the total intensity of magnetic scattering as a function of magnetic field, similar sector boxes were chosen as shown in Fig. S9. The resulting intensity evolution is shown in (H). |q| corresponding to the maximum of intensity was estimated for each of the detector images shown and is plotted in (I). It also shows the equivalent pitch length in real space for these incommensurately modulated phases. The dashed line in magenta signifies the height of each octahedron, as estimated from the W-H analysis presented in Fig. 2E of the main text.

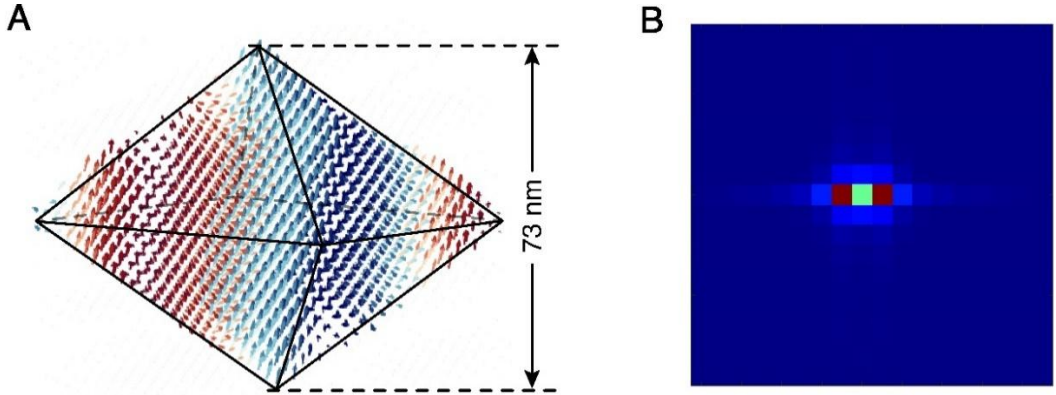

**Fig. S11.** (A) A single period of magnetic helix embedded into 73 nm nanoparticle. (B) Fast-Fourier transform of the *z*-component of the magnetization projected on the (*yz*) plane, showing characteristic "two-spot" pattern.

**Micromagnetic simulations.** The magnetization *M* component along the direction of the applied field was extracted from the simulations for both *B* ∥ [001] (along the particle height, as seen in Fig. 7A) and *B* ∥ [111] (along the particle facet, in Fig. 7B). Metamagnetic transitions seen in both *M* (*B*) and d*M*/d*B* graphs illustrate the evolution of magnetic textures shown in Fig. 6.

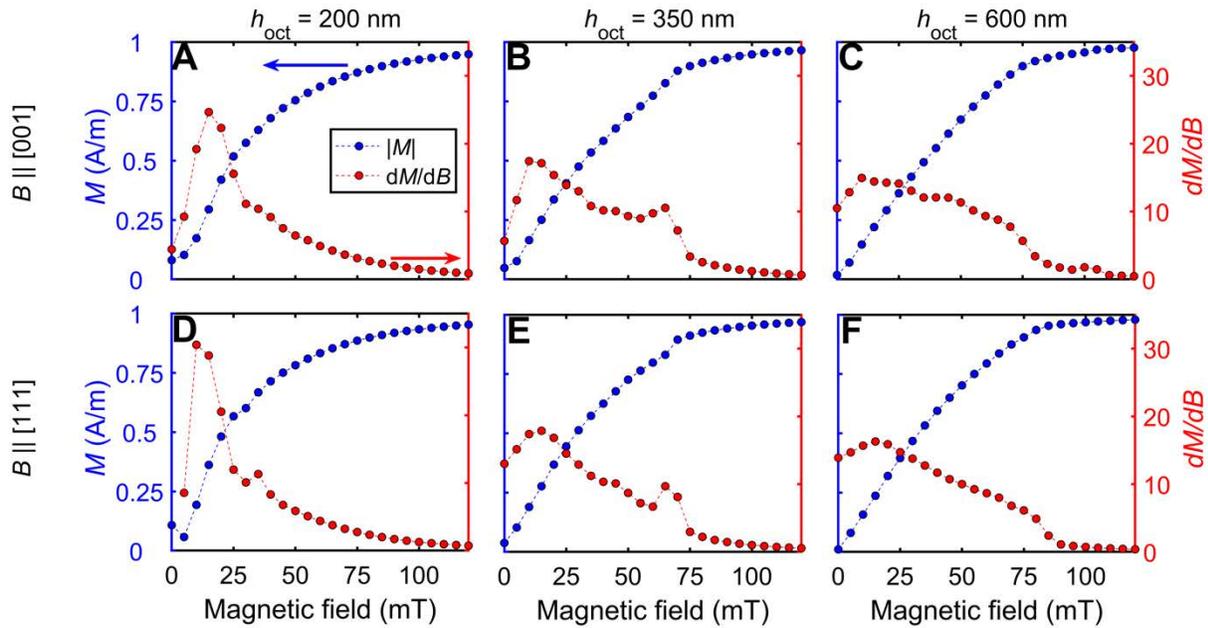

**Fig. S12.** For the three sizes shown in Fig. 6, evolution of the total magnetization as a function of magnetic field, as obtained from micromagnetic simulations. Total magnetization refers to the norm of the three spatial components. (A)-(C) represent when the magnetic field was applied along the top vertex of the octahedron ([100]), whereas (D)-(F) represent those along one of the facets ([111]). d*M*/d*B* clearly shows the metamagnetic-type transitions in individual nanoparticles (for example, panel B).

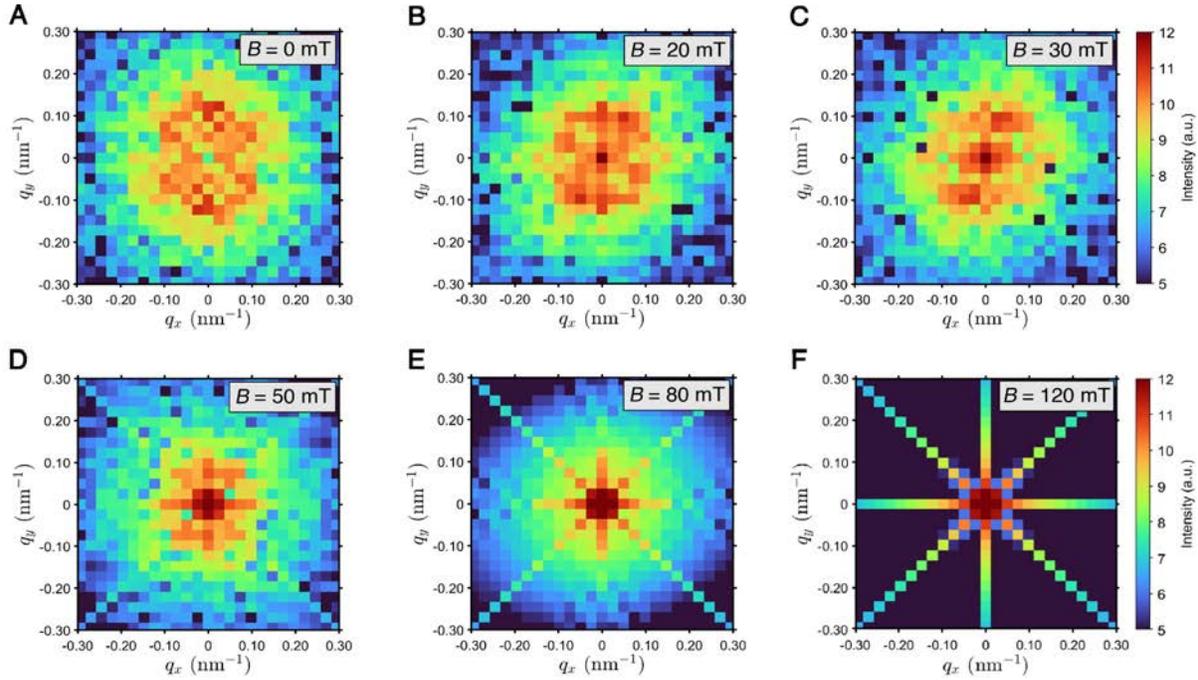

**Fig. S13.** Magnetic field evolution of the Fast Fourier Transformation of the *z*-component of the magnetization, $M_z \parallel \mathbf{B}$ projected onto the $(q_x, q_y)$ plane for the 512 nm particle.

To distinguish between the possible signatures of new three-dimensional (3D) magnetic textures we have analyzed results of the micromagnetic simulations of 512 nm particles via Fast-Fourier Transformation (FFT) of the magnetization component $M_z \parallel \mathbf{B}$ projected onto the $(q_x, q_y)$ plane. The simulation results are shown in Figure S13 as a function of the magnitude of the applied field covering the characteristic phases (see Fig. 7 of the main text): helical (B = 0 mT), helical + meron (20 mT), meron (30 mT), skyrmion + chiral bobber (50 mT), single skyrmion (80 mT) and, finally, induced ferromagnetic (120 mT). The FFTs show maxima centered at $|q|\sim 0.1$ nm$^{-1}$ in all phases hosting multiple magnetization kinks (or magnetic quasiparticles) which corresponds well to the characteristic wavevector of magnetic modulations in $Cu_2OSeO_3$. These maxima are superimposed onto the broad halo originated from the magnetic form-factor scattering from the single magnetic kinks, especially pronounced in the case of the isolated skyrmion (80 mT). The cross-shaped intensity distribution at 120 mT represents the canted moments at the edges of the octahedron in the field-induced ferromagnetic state. The accurate fitting of the small-angle scattering profiles and their comparison to micromagnetic models would be a promising way to extract the 3D magnetic textures such as merons, vortices and chiral bobbers. However, we note that the present calculation is done for the nanoparticle and does not account for the powder averaging of the scattering pattern and for the magnetization distribution within separated particles, which would smear out the characteristic features of the form factor (see the detailed analysis of the particle size distribution on magnetic SANS in the recent preprint[2]). Therefore, unique identification of these phases are more likely to be completed via real-space probes, or scattering experiments that can be performed on the single-crystal-particle level, such as small-angle electron diffraction or resonant x-ray scattering.

1. P. Stadelmann. (EPFL, Lausanne, 2014-2020), pp. *Electron microscopy simulation software simulates electron diffraction, high-resolution TEM and STEM images, and 3-D images of the crystal*.

2. M. P. Adams, A. Michels, and H. Kachkachi, arXiv preprint arXiv:2205.07552 (2022).